\documentclass[aps,reprint,prl,showpacs,superscriptaddress]{revtex4-2}

\usepackage{amsmath}
\usepackage{mathrsfs,ulem,soul,bm}
\usepackage{siunitx}
\usepackage{graphicx}

\begin{document}

\title{Disorder-induced stress-flow misalignment in soft glassy materials revealed using multi-directional shear}

\author{Frédéric Blanc}
\affiliation{Institut de Physique de Nice, UMR7010, CNRS, Université Côte d'Azur, 17 rue Julien Lauprêtre, 06200 Nice, France}
\author{Guillaume Ovarlez}
\affiliation{Univ. Bordeaux, CNRS, Syensqo, LOF, UMR 5258, F-33600 Pessac, France}
\author{Adam Trigui}
\author{Kirsten Martens}
\author{Romain Mari}
\affiliation{Univ. Grenoble-Alpes, CNRS, LIPhy, 38000 Grenoble, France}

\begin{abstract}
Controlling the mechanical response of soft glassy materials--such as emulsions, foams, and colloidal suspensions--is key for many industrial processes. While their steady-state flow behavior is reasonably well understood, their response to complex flow histories, as encountered in operations like pumping or mixing, remains poorly known. Using a custom multi-axis shear apparatus that enables arbitrary changes in flow direction, we investigate how shear history influences the mechanical behavior of a model soft glassy system. We uncover a transient shear response orthogonal to the applied shear direction, together with an anisotropic yield surface. These effects point to an underlying anisotropic distribution of internal stresses imprinted by previous deformation. To rationalize this behavior, we use a mesoscopic elasto-plastic model, demonstrating that local mechanical disorder governs the emergence of macroscopic stress–flow misalignment. Our findings offer a new route to experimentally probe the distribution of local yield stresses in soft glassy materials.
\end{abstract}


\maketitle




Materials such as gels, foams, cement pastes, and biological tissues have the intriguing ability to transition from a solid-like state to a fluid-like state upon application of stress~\cite{coussot2014yield,bonnYieldStressMaterials2017}. From a mechanical perspective, they are classified as yield-stress fluids (YSFs): below a critical yield stress, they behave as elastic solids; above it, they flow like viscous fluids~\cite{balmforth2014yielding}. Physically, these materials exhibit properties similar to glasses, including disorder and arrested dynamics, which justifies their classification as soft glassy materials~\cite{Hunter_2012,nicolas2018deformation, divoux2024ductile}. Understanding their rheological behavior is essential for improving formulations and processes in industrial applications and for advancing our knowledge of glass physics.

In macroscopic rheology, the focus has long been on viscoplasticity; the elastoplastic behavior of YSFs has been largely disregarded. 
Their steady flow behavior is typically modeled as a Herschel-Bulkley behavior, with an assumed viscous tensorial form where the deviatoric stress and the strain-rate tensors are codirectional~\cite{saramito2007new,saramitoNewElastoviscoplasticModel2009,ovarlez2010three,balmforth2014yielding, kamani2021unification}. Elasticity however also plays a crucial role in many fluid mechanics problems~\cite{fraggedakis2016yielding1,fraggedakis2016yielding2}. 
The complete elastoviscoplastic behavior of YSFs has been first described as a perfect isotropic elastoplastic behavior~\cite{saramito2007new,saramitoNewElastoviscoplasticModel2009, kamani2021unification}, providing a fair prediction of fluid mechanics problems~\cite{cheddadiUnderstandingPredictingViscous2011,fraggedakis2016yielding1, kamani2024brittle, agrawal2025features}.

The link between the steady-state rheological properties of these materials and their structure and interactions at the microscopic scale is reasonably well understood~\cite{bonnYieldStressMaterials2017,braderFirstPrinciplesConstitutiveEquation2008,braderGlassRheologyModecoupling2009,cunyDynamicsMicrostructureAnisotropy2022}, for various materials such as microgels and emulsions~\cite{sethMicromechanicalModelPredict2011}, colloidal glasses~\cite{Hunter_2012}, or foams~\cite{cohen2013flow}. The macroscopic behavior of YSFs is more generally well accounted for by mesoscopic models, which consider that their flow is governed by the collective dynamics of the microstructural elements of the material~\cite{sollich1997rheology,hebraud1998mode,langer2008shear}, be they composed of droplets, particles, or bubbles. 
Mesoscopic elastoplastic models conceptualize the material as a collection of mesoscopic blocks 
that switch between elastic behavior and plastic relaxation when loaded above a threshold~\cite{nicolas2018deformation}. 
Stress redistribution occurs as plastic relaxation events take place in the system, which eventually cause avalanches near the yielding transition. 
These models and their continuum counterparts \cite{bocquet2009kinetic, moorcroft2013criteria, benzi2016cooperativity} have historically been restricted to scalar, shear-stress-only formulations.
This simplification was shown to be adequate when compared with early tensorial implementations under unidirectional simple shear protocols~\cite{nicolas2014universal} and simple unidirectional flow experiments \cite{goyon2008spatial, benzi2019unified}. Later developments incorporated local disorder in yield stresses into tensorial models to capture plasticity bursts in the jammed phase~\cite{budrikisUniversalFeaturesAmorphous2017} under various deformation protocols, and further reinforced the validity of scalar model predictions, particularly regarding universal avalanche statistics in the stationary state at small rates of strain.
Within these simplified scalar descriptions, it has been demonstrated that both the residual local stresses and the spatial heterogeneity of local yield stresses play a crucial role in determining the transient response under unidirectional simple shear~\cite{liu2018creep, ozawa2018random, popovic2018elastoplastic, patinet2020origin, ruan2022predicting}. 
Efforts to characterize the initial stress and yield stress fields~\cite{puosi2015probing, patinet2016connecting} have enabled semi-quantitative predictions of the transient dynamics, particularly in the creep and start-up regimes~\cite{liuElastoplasticApproachBased2021, castellanos2021insights}.

The dependence of the elastoplastic response of yield stress fluids (YSFs) on the history of deformation is more complex and less understood than the steady-state behavior. 
In this context, a remarkable phenomenon is the Bauschinger effect: after steady plastic flow, these materials exhibit a softer stress response under reverse loading compared to reloading
\cite{vincent2011work,dimitriou2013describing,bhattacharjee2015stress,dimitriou2019canonical,patinet2020origin,deboeufMechanismStrainHardening2022}. 
Its importance has been highlighted by the integration of kinematic hardening into a continuum model describing the elasto-visco-plastic behavior of YSFs, which is able to capture the shear stress response under large-amplitude oscillatory shear~\cite{dimitriou2013describing,dimitriou2019canonical}.
Anisotropic hardening in small-amplitude oscillatory shear has also been recently observed~\cite{beyerShearinducedMechanicalAnisotropy2025}.

Strain hardening and Bauschinger effects are observed in various plastic materials, including crystals~\cite{seegerWorkhardeningWorksofteningFacecentred1957,hirsch1964theory}, amorphous solids~\cite{hoy2008strain,amesThermomechanicallyCoupledTheory2009,anandThermomechanicallyCoupledTheory2009,pan2020strain,patinet2020origin}, composite materials~\cite{li2001tensile}, and a wide range of gels~\cite{weigandt2011situ,groot1996molecular,an2012direct,laurati2014plastic}. 
At the microscopic level, these effects arise from mechanisms such as defect interactions triggering local plastic events in crystalline solids~\cite{Buckley56,Asaro75,Frank80} or the development of microstructural anisotropies in suspensions of particles in YSFs~\cite{deboeufMechanismStrainHardening2022}. 
At the mesoscopic scale, simulations of amorphous solids~\cite{Karmakar10,patinet2020origin}  and colloidal glasses~\cite{ederaTuningResidualStress2025} reveal that plastic deformation induces anisotropy in the internal stress distribution, leading to an asymmetric macroscopic response depending on the direction of applied deformation. 
These progresses in understanding the physical origin of scalar strain hardening and Bauschinger effects beg for an in-depth investigation of the tensorial aspects of these phenomena.

To address this issue, we use an original two-dimensional shear experiment to study the 2D elasto-plastic response of a model YSF under complex flow histories. 
Whereas standard studies of the Bauschinger effect provide 1D information (shear stress in the flow direction) for 1D strain histories (response to shear reversal), our approach enriches significantly the description with 2D information (shear stresses both parallel and orthogonal to flow) for 2D strain histories (response to any change in flow direction). 
Our experiment reveals an anisotropic Bauschinger effect, and the associated development and relaxation of an unexpected shear component orthogonal to the shear direction, which we demonstrate are signatures of internal stress relaxation. 
The overall transient behavior of stresses is  captured by a tensorial generalization of a mean-field mesoscopic elasto-plastic model, revealing the role of the distribution of local yield stresses within the material.

\section*{\bf Experimental results}

\begin{figure}
    \centering
\includegraphics[width=\linewidth]{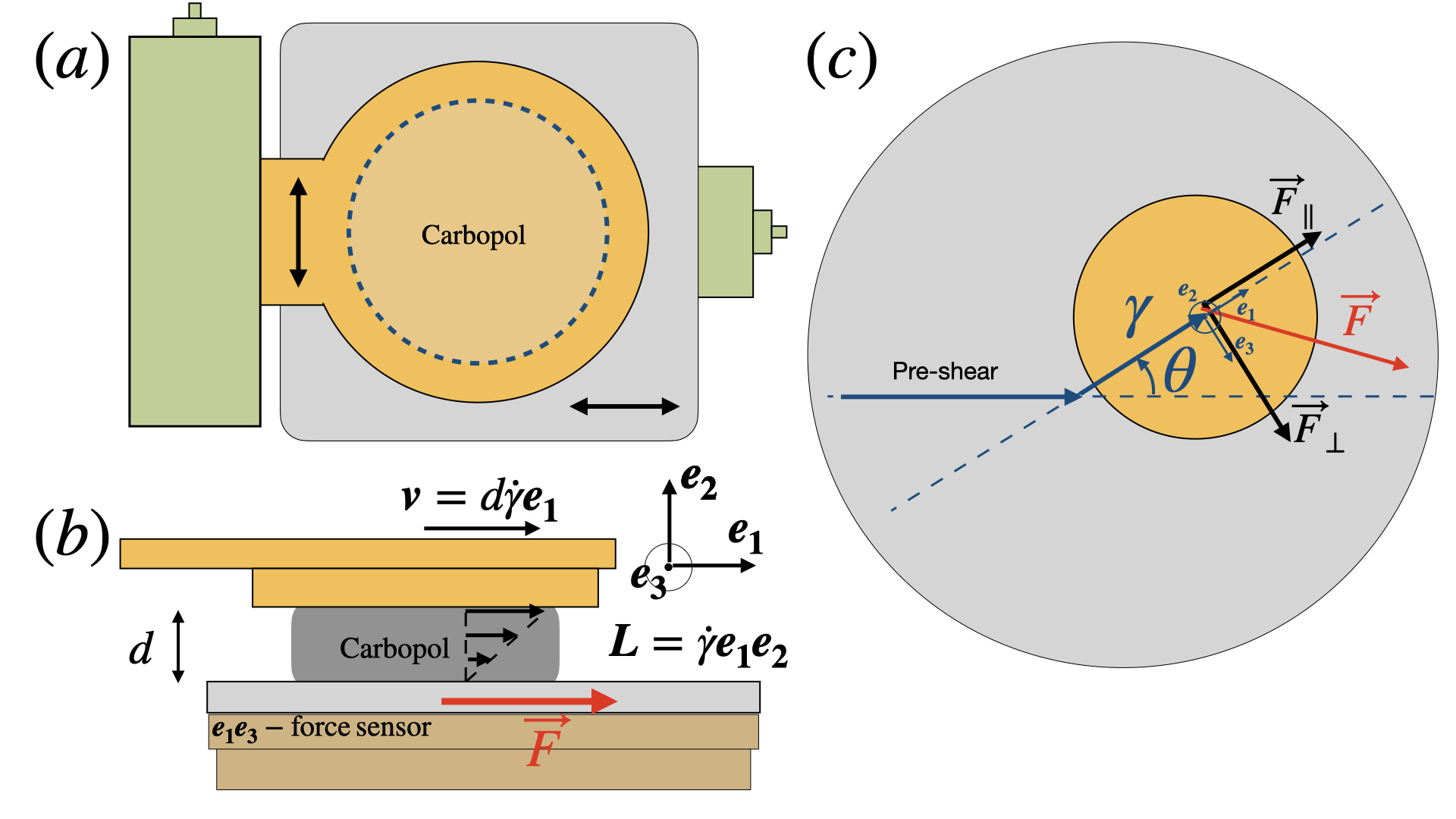}
    \caption{\footnotesize Sketch of the 2D rheometer (2dR). (a) View in the flow-vorticity plane $(\bm{e}_1,\bm{e}_3)$. Two translation stages (light green) independently move in orthogonal directions an upper plate (orange) and a lower plate (light gray), between which the material is sheared. (b)  View in the flow-gradient plane $(\bm{e}_1,\bm{e}_2)$. A force sensor measures both the amplitude and the direction of the shear force $\bm{F}$. (c) View in the flow-vorticity plane of the trajectory (thick dark blue line) of the top plate relative to the bottom plate when flow direction is changed by an angle $\theta$. A transient force $\bm{F} = F_{\parallel}\bm{e}_1 + F_{\perp}\bm{e}_3$ is recorded, with respect to the post rotation strain ($\gamma$), after rotation of the flow direction.}
    \label{fig:CR_setup}
\end{figure}

We use an original in-house developed 2D rheometer (2dR)~\cite{blancRheologyDenseSuspensions2023}. In the 2dR, the YSF is sheared between two parallel plates moving independently and orthogonally to each other, and the resulting shear force acting on the plates is measured (see Materials and Methods for details). A simple shear flow of shear rate $\dot\gamma$ can be enforced in any direction $\bm{e}_1$ in the horizontal plane with a velocity gradient in the vertical direction $\bm{e}_2$ (Fig.~\ref{fig:CR_setup}). The resulting shear force $\bm{F} = F_{\parallel} \bm{e}_1 + F_{\perp} \bm{e}_3$ (with $\bm{e}_3 = \bm{e}_1 \times \bm{e}_2$ the vorticity direction) then provides two shear stress components $\Sigma_\parallel= F_\parallel / S$ and $\Sigma_\perp = F_\perp / S$, where $S$ is the contact area between the YSF and the plates. The main originality of this setup lies in (i) the ability to arbitrarily change the flow direction in the horizontal plane and thus go beyond standard shear reversal experiments, and (ii) the characterization of the shear stress $\Sigma_\perp$ in a direction orthogonal to flow (vorticity direction) in addition to the standard $\Sigma_\parallel$ measurement.

Here, we use the 2dR to examine the elastoplastic response of a Carbopol gel (see Materials and Methods) when subjected to an arbitrary change in shear direction within the horizontal plane. Carbopol gels are dense suspensions of microgel particles, widely used as model YSFs~\cite{piau2007carbopol,balmforth2014yielding,coussot2014yield}. They exhibit a marked Bauschinger effect~\cite{mahaut2008yield,dimitriou2013describing,deboeufMechanismStrainHardening2022}.

In all experiments, we shear the material at a low constant shear rate $\dot\gamma=\SI{3e-2}{\per\second}$ to minimize viscous effects. The material is first presheared until steady plastic flow is observed, characterized by a steady shear stress $\Sigma^\mathrm{ss}_{\parallel}$ approximately \SI{18}{\percent} higher than the measured dynamic yield stress $\Sigma^\mathrm{d}_y = \SI{120}{\pascal}$ of the Herschel-Bulkley flow curve fit (see Materials and methods).
In steady shear, we measure $\Sigma^\mathrm{ss}_{\perp}=0$, as expected. 
The microgel is then unloaded at the same shear rate until the macroscopic stress is fully relaxed to zero. 
Starting from rest, the 2D elastoplastic response of the material is then studied by shearing it in directions at various angles $\theta$ with respect to the preshear direction. In the following, we note  $\Sigma^\theta_{\parallel,\perp}$ the stress components observed after rotation of the flow direction by an angle $\theta$. 

\begin{figure}[h!]
    \centering
    \includegraphics[width=\linewidth]{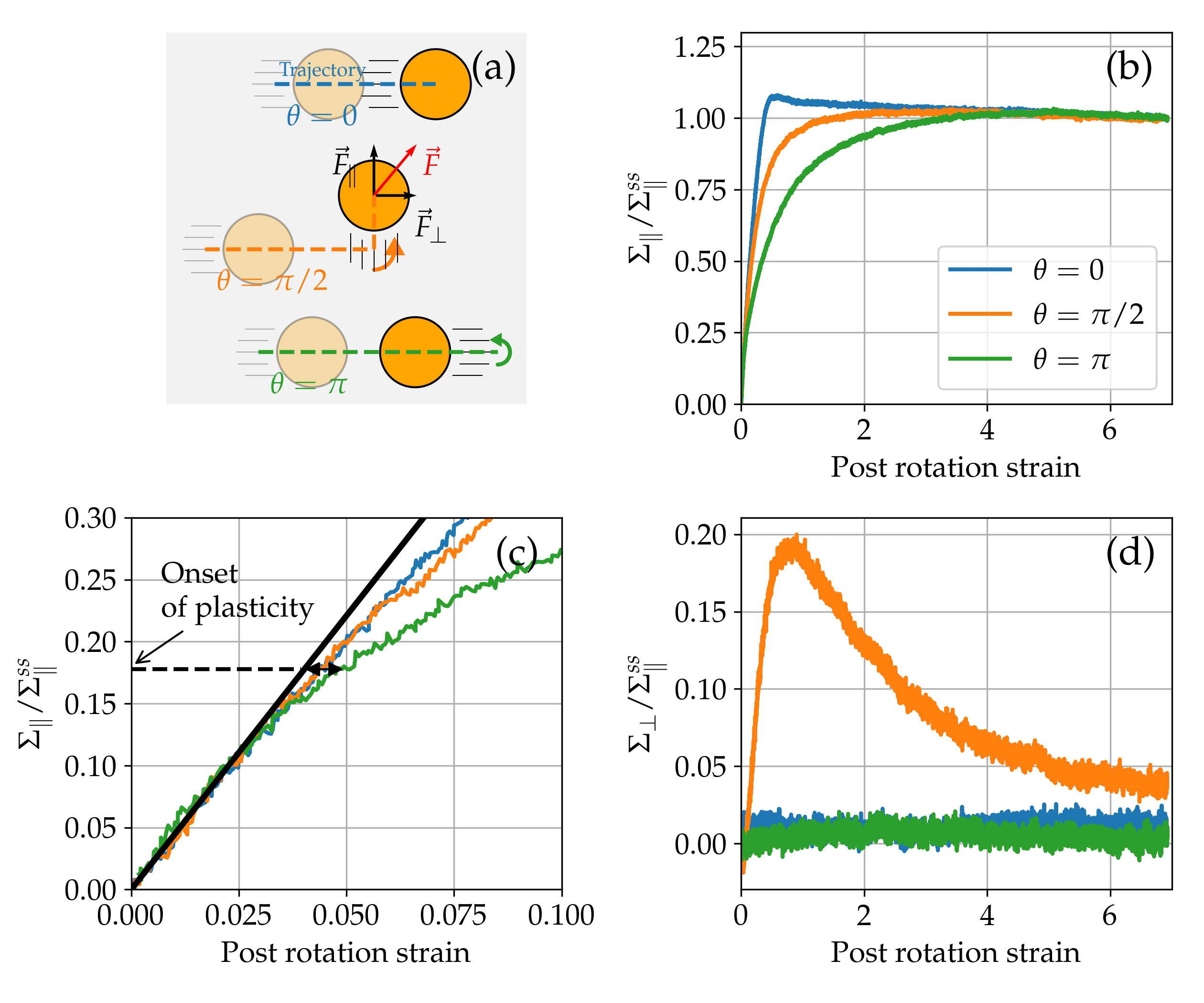}
    \caption{\footnotesize Two-dimensional response of a Carbopol gel to a change of the shear direction in the 2D rheometer of Fig.~\ref{fig:CR_setup}; flow is rotated by an angle $\theta$ with respect to the preshear direction. (a) The blue, orange, and green hashed lines represent three specific trajectories. (b) Dimensionless shear stress $\Sigma^{\theta}_\parallel/\Sigma^\mathrm{ss}_\parallel$ aligned with the shear direction, for $\theta=0$, $\pi/2$, and $\pi$ (shear reversal). (c) Zoom-in on figure (b). The thick black line represents the linear fit for post-rotation strain < 0.03. The deviation by 1\% strain from this linear behavior (indicated by the dashed line) defines the stress $\Sigma_\mathrm{p}^\theta$ at the onset of plasticity for a given angle $\theta$. (d) Dimensionless shear stress $\Sigma^{\theta}_\perp/\Sigma^\mathrm{ss}_\parallel$ orthogonal to the shear direction for $\theta=0$, $\pi/2$, and $\pi$. Same colors as in (b).  }\label{fig:CR_carbopol}
\end{figure}

We first examine the standard cases $\theta = 0$ (reloading) and $\theta= \pi$ (shear reversal), as shown in Fig.~\ref{fig:CR_carbopol}a (see Materials and Methods for conventional rheometric measurements).
The loading curves are displayed in Figs.~\ref{fig:CR_carbopol}b--d.
During reloading (blue curves), the material exhibits an almost perfect elastoplastic response with applied strain, characterized by an approximately linear increase of $\Sigma^0_\parallel$
followed by a steady-state plateau at $\Sigma^0_\parallel=\Sigma^\mathrm{ss}_\parallel$ (Fig.~\ref{fig:CR_carbopol}b). 
A small overshoot is observed, but we do not discuss it further here (see, e.g., \cite{vasisht2022residual, ederaTuningResidualStress2025} for a discussion). These features indicate that the material underwent strain hardening in the initial shear.
In contrast, the material appears much softer with a ductile behaviour and small yield stress upon shear reversal ($\theta=\pi$, green curves). During the shear the material undergoes again strain hardening, the shear stress $\Sigma^\pi_\parallel$ increases more and more slowly with strain and reaches a steady state only at a strain of approximately 4. 
This asymmetric response is consistent with the literature~\cite{mahaut2008yield,dimitriou2013describing,deboeufMechanismStrainHardening2022}.
In both cases ($\theta=0$ and~$\pi$), $\Sigma^\theta_\perp$ remains zero throughout the experiment (Fig.~\ref{fig:CR_carbopol}d). 

Now we turn to a flow-direction rotation of angle $\theta = \pi/2$ (orange curves in Fig.~\ref{fig:CR_carbopol}).
The material also exhibits a softened response and exhibits strain hardening, although less marked than in shear reversal (Fig.~\ref{fig:CR_carbopol}b). 
Intriguingly, we observe a previously unreported nonlinear phenomenon, as a non-zero shear stress component $\Sigma^{\pi/2}_\perp$ emerges orthogonal to the flow direction (Fig.~\ref{fig:CR_carbopol}d).
The value of $\Sigma^{\pi/2}_\perp$ increases to a maximum of approximately $0.2\Sigma^\mathrm{ss}_\parallel$ at a strain of about 1 and then slowly decreases towards zero with further strain. 
A non-zero $\Sigma_\perp$ indicates that the force required to make the material flow is transiently misaligned with the motion direction, or, equivalently, that the stress tensor is not co-directional with the applied strain-rate tensor.
Interestingly, the transient behavior of $\Sigma^{\pi/2}_\perp$ persists much longer than that of $\Sigma^{\pi/2}_\parallel$. 
While $\Sigma^{\pi/2}_\parallel$ appears to reach a steady response at a strain of about 2, $\Sigma^{\pi/2}_\perp$ decreases slowly with strain throughout the experiment and has not yet reached a steady state at a strain of 7. 
This indicates a continued slow internal dynamics with no observable impact on $\Sigma^{\pi/2}_\parallel$.

\begin{figure*}
    \centering
    \includegraphics[width=0.95\linewidth]{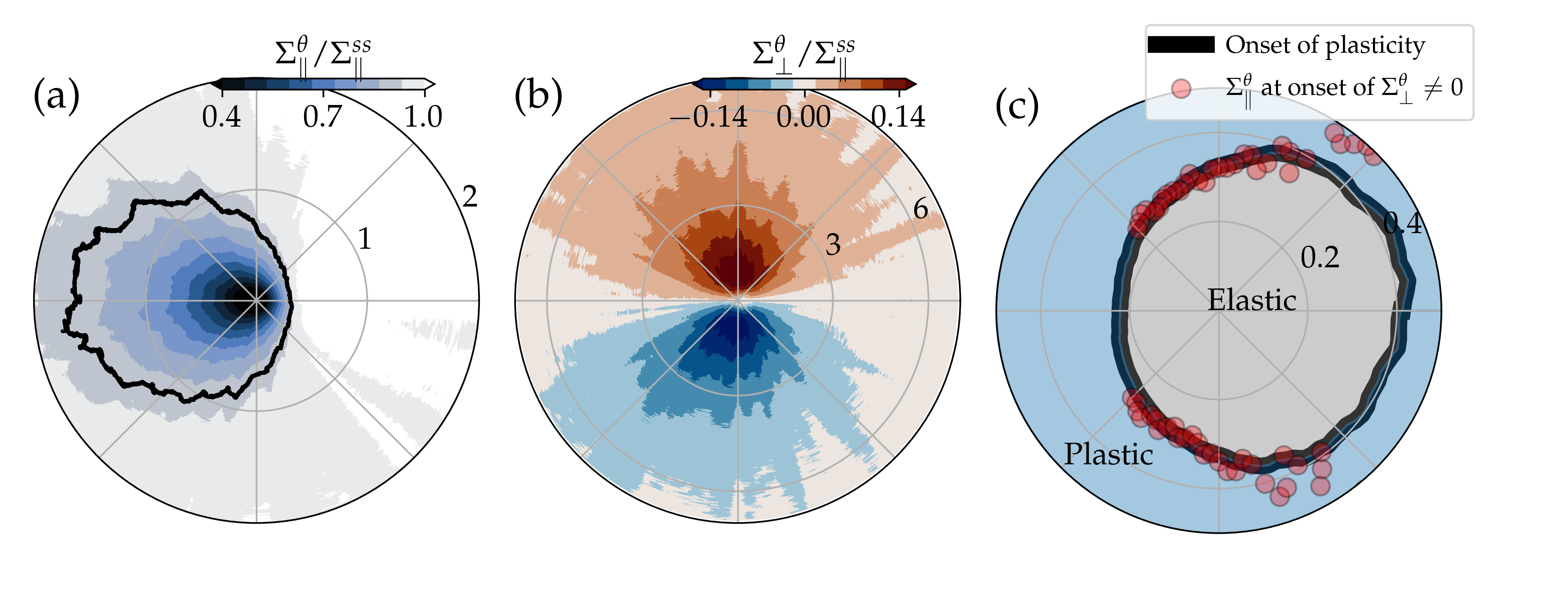}
    \caption{
    (a) and (b) : Two-dimensional response of a Carbopol gel to a change in flow direction in the 2D rheometer shown in Fig.~\ref{fig:CR_setup}. Flow is rotated by an angle $\theta$ relative to the preshear direction. The center of each plot corresponds to the rest state, obtained after preshear and unloading. The radial coordinate represents the post-rotation strain $\gamma$, and the angular coordinate corresponds to the rotation angle $\theta$. (a) Dimensionless shear stress $\Sigma^{\theta}_\parallel/\Sigma^\mathrm{ss}_\parallel$, aligned with the new shear direction. The black contour marks the level $\Sigma^{\theta}_\parallel = 0.9\Sigma^\mathrm{ss}_\parallel$. (b) Dimensionless shear stress $\Sigma^{\theta}_\perp/\Sigma^\mathrm{ss}_\parallel$, orthogonal to the shear direction. (c) Yield surface of the carbopol gel (black line), shown in polar coordinates: the onset of plasticity 
    $\Sigma^\theta_\mathrm{p}$ is normalized by the dynamic yield stress $\Sigma^\mathrm{d}_\mathrm{y}$ and plotted as a function of the flow-direction rotation angle $\theta$.
    The black line separates the elastic region (light gray) from the plastic region (light blue). In red symbols, we plot the value of the parallel stress $\Sigma^{\theta}_\parallel$ at the onset of $\Sigma^\theta_\perp\neq 0$.}
    \label{fig:contour_F12_F32_exp}
\end{figure*}

This behavior persists across all rotation angles, as illustrated in Fig.~\ref{fig:contour_F12_F32_exp}.
The full range of flow-direction rotations, $\theta \in [-\pi, \pi]$, is explored by repeating the experiments at $3^\circ$ increments.
The responses of $\Sigma^{\theta}_\parallel$ and $\Sigma^{\theta}_\perp$ are displayed as polar color maps: the radial coordinate represents the post-rotation strain $\gamma$, the angular coordinate corresponds to $\theta$, and the color scale encodes the amplitudes of $\Sigma^{\theta}_\parallel(\gamma)$ (Fig.~\ref{fig:contour_F12_F32_exp}a) and $\Sigma^{\theta}_\perp(\gamma)$ (Fig.~\ref{fig:contour_F12_F32_exp}b).
The center of each plot corresponds to the rest state, obtained after preshear and unloading.
From these maps emerge the key experimental findings of this study:
(i) an orthogonal stress component $\Sigma^{\theta}_\perp$ develops for all $\theta \neq 0$ and $\theta \neq \pi$, and
(ii) the transient response extends over significantly larger strains in $\Sigma^{\theta}_\perp$ than in $\Sigma^{\theta}_\parallel$.
Moreover, the data reveal that the yield stress decreases with increasing $|\theta|$, reaching a minimum at $\theta = \pi$, meaning that the strain hardening intensifies under re-shear.
The black contour line in Fig.~\ref{fig:contour_F12_F32_exp}a, which corresponds to $0.9\Sigma_{\parallel}^\mathrm{ss}$, highlights the anisotropic character of the Bauschinger effect.

We now turn to the yield surface of the material.
As illustrated in Fig.~\ref{fig:CR_carbopol}c,
we define the onset of plasticity $\Sigma_\mathrm{p}^{\theta}$ as the stress at which the stress-strain curve deviates by \SI{1}{\percent} from initial linearity. $\Sigma_\mathrm{p}^{\theta}$ values are shown as a black line in a polar plot (Fig.~\ref{fig:contour_F12_F32_exp}c), where the radial coordinate now represents $\Sigma_\mathrm\parallel^{\theta}$. The resulting yield surface is markedly anisotropic. By contrast, the linear elastic behavior does not show any significant anisotropy (see Fig.~\ref{fig:CR_carbopol}c for $\gamma\lesssim0.02$).
Consistently, the anisotropy of  Fig.~\ref{fig:contour_F12_F32_exp}c contrasts with that observed in Fig.~\ref{fig:contour_F12_F32_exp}a.
This inversion arises from the fact that plasticity develops more gradually at large values of $|\theta|$ than at small values.

\section*{\bf Qualitative interpretation}

\begin{figure}
    \centering
    \includegraphics[width=\linewidth]{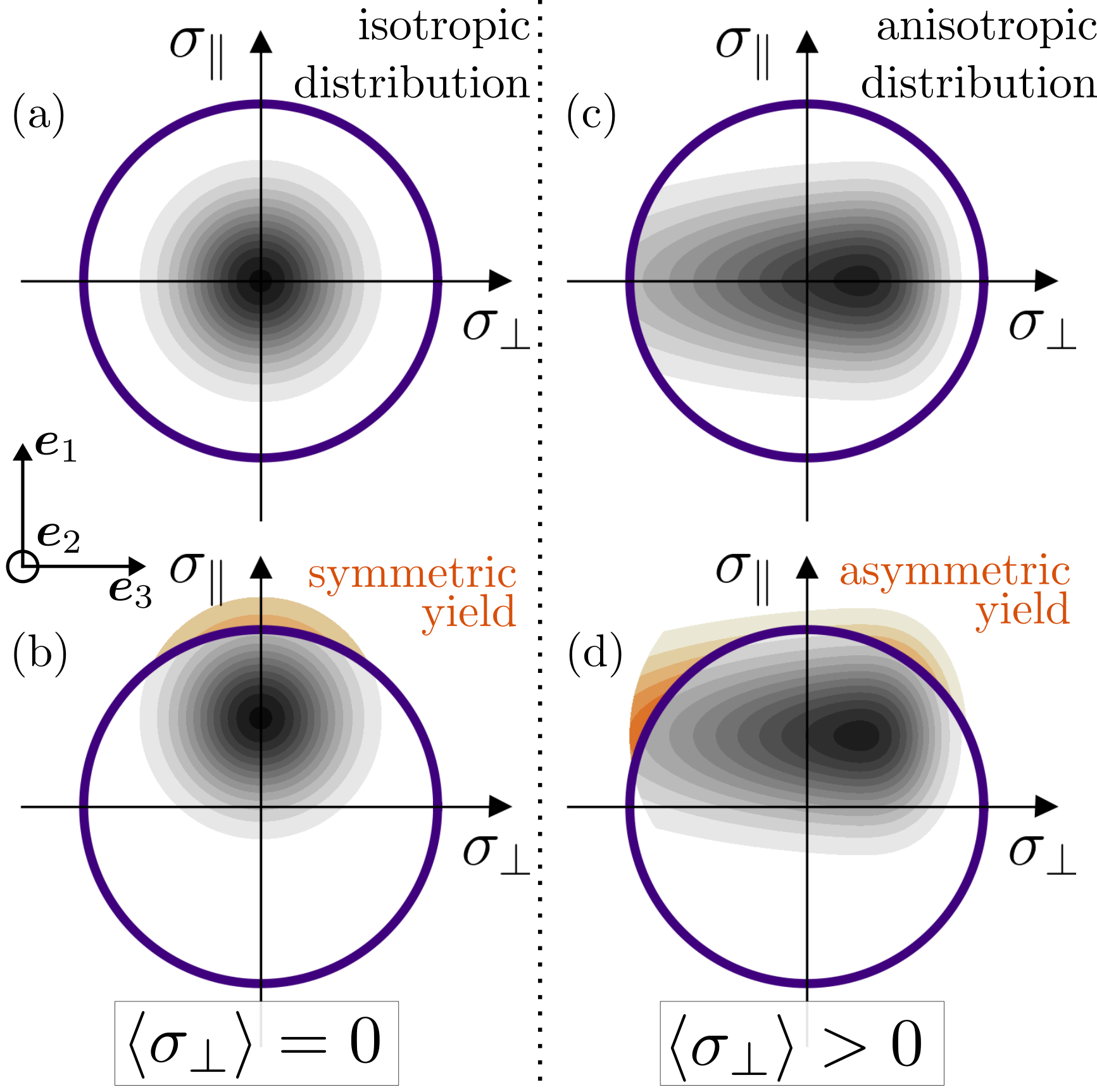}
    \caption{Mechanism: yielding in presence of an anisotropic internal stress distribution induces finite orthogonal stresses. 
    (a) Isotropic internal stress distribution at rest. (b) Same distribution, but driven to yield. Yielding is symmetric with respect to the $\sigma_\parallel$ axis, keeping the average orthogonal stress $\langle \sigma_\perp \rangle = 0$.
    (c) Anisotropic internal stress distribution from previous deformation. (d) Same distribution, but driven to yield. 
    Yielding is asymmetric with respect to the $\sigma_\parallel$ axis, leading to $\langle \sigma_\perp \rangle > 0$.}
    \label{fig:mechanism}
\end{figure}

We argue that the anisotropic macroscopic yield surface and the nonzero orthogonal shear component after the rotation of the flow direction are both signatures of an anisotropic internal stress distribution imprinted by the preshear.
At a mesoscopic scale, the local  stress components  are fluctuating quantities.
At rest, the distribution of these internal stresses lies within the local 
yield surface and is such that its average -- the macroscopic stress -- vanishes. 
In Fig.~\ref{fig:mechanism}, we sketch the consequence of a shear in an arbitrary direction $\bm{e}_\parallel$, which defines local stress components $\sigma_\parallel$ and $\sigma_\perp$, in a simplified scenario with constant local yield stress.
For an isotropic distribution of internal stresses (Fig.~\ref{fig:mechanism}a), shear advects the distribution out of the yield surface in the $\bm{e}_\parallel$ direction, leading to a buildup of the macroscopic parallel stress $\Sigma_\parallel = \langle \sigma_\parallel \rangle$ (Fig.~\ref{fig:mechanism}b).
Yield events that relax the stress occur in a symmetric manner around $\bm{e}_\parallel$, such that 
the macroscopic orthogonal stress $\Sigma_\perp = \langle \sigma_\perp \rangle$ remains zero.
Shear in any other direction would result in the same onset of plasticity.
By contrast, for an anisotropic distribution (Fig.~\ref{fig:mechanism}c), shear leads to an asymmetric yield, which gives rise to a finite $\Sigma_\perp$ (Fig.~\ref{fig:mechanism}d).
Moreover, the onset of plasticity will be direction dependent.
Crucially, because the finite orthogonal stress results from anisotropic plasticity, we expect that the onset of plasticity is concurrent to the rise of $\Sigma_\perp$. To test this idea, we measure the value of the parallel stress at the point where the orthogonal shear stress $\Sigma^\theta_\perp$ starts evolving (see Supporting Information Fig.~S1a). Given the small values of $\Sigma^\theta_\perp$ for $\theta$ close to $0$ and $\pi$, this analysis is restricted to angles $|\theta| \in [\pi/4, 3\pi/4]$.
The results, displayed in Fig.~\ref{fig:contour_F12_F32_exp}c, show remarkable agreement with the yield surface, confirming the critical role of $\Sigma_\perp$ as a sensitive indicator of plasticity.

In order to model quantitatively the behaviors of $\Sigma_\perp$ and of the macroscopic yield surface, we need an accurate representation of the internal stress distribution.
In a realistic scenario, the local yield is broadly distributed, which will in turn strongly affect the internal stress distribution.
This is the purpose of the next section.

\section*{\bf Minimal mean-field model}

The phenomenology observed experimentally requires accounting for two key ingredients: an asymmetric distribution of internal stresses and disorder in the local yield stress field. 
A realistic model must therefore incorporate both features. 
In this section, we aim to construct a minimal mean-field description that includes these essential aspects. 
We consider 
a yield stress fluid subject to an arbitrary shear in an $(x,z)$ plane, with direction $y$ as the shear gradient direction.
Following the spirit of the Hébraud-Lequeux model~\cite{hebraud1998mode}, 
we divide our system into $N$ mesoscopic regions, and we attribute to each region a local two-component stress $\bm{\sigma} = (\sigma_{xy}, \sigma_{zy})$ and a local yield stress $\sigma^\mathrm{c}$. 
In a mean-field approximation, the dynamics of these mesoscopic regions can be described with a Langevin equation (here for site $i$)
\begin{equation}\label{eq:langevin}
    \partial_t\bm{\sigma}_i(t) = G \bm{E} + \sqrt{2 \alpha \Gamma_i(t)}\bm{\xi}_i(t)
\end{equation}
with an applied flow $\bm{E} = (E_{xy}, E_{zy})$, an uncorrelated noise $\langle \bm{\xi}_i(t) \bm{\xi}_j(t') \rangle = \delta_{ij}\delta(t-t')\bm{1}$, with $\bm{1}$ the identity matrix, and $\alpha>0$ a model parameter.
The mechanical noise amplitude depends on the total plastic activity rate as
\begin{equation}
     \Gamma_i(t) = \frac{1}{(N-1) \tau} \sum_{j\neq i} \Theta(\lvert \bm{\sigma}_j \rvert - \sigma^c_j) \, .
\end{equation}
At each time step, sites above their yield stress are susceptible to yield following a local von Mises yield criterion
\begin{equation}\label{eq:langevin_yield}
    \bm{\sigma}_i \xrightarrow[]{\tau^{-1}} \bm{\sigma}^\mathrm{r} \text{ if } \lvert \bm{\sigma}_i \rvert \geq \sigma^\mathrm{c} _i\, .  
\end{equation}
Upon yielding, the site is assigned a new local yield threshold, randomly drawn from a yield stress distribution $\rho(\sigma^c)$, , and its stress is reset to a relaxed value $\bm{\sigma}^\mathrm{r}$, drawn from a distribution $\rho_\mathrm{r}(\sigma^\mathrm{r})$.
The yield stress distribution will turn out to be key to control the behavior under rotation of the flow direction.
In the limit of large number of representative sites Eqs.~\ref{eq:langevin} and~\ref{eq:langevin_yield} give rise to a distribution $\mathcal{P}(\bm{\sigma}, \sigma_\mathrm{c}, t)$ which follows a 2-component Hébraud-Lequeux equation (see Materials and Methods).
Finally, the macroscopic stress of the material is given by $\bm{\Sigma} = \langle \bm{\sigma}\rangle$.
For $\alpha<1/4$, the model exhibits a finite yield stress and a Herschel-Bulkley flow curve. 

\section*{\bf Model predictions}

\begin{figure}
    \centering
        \includegraphics[width=\columnwidth]{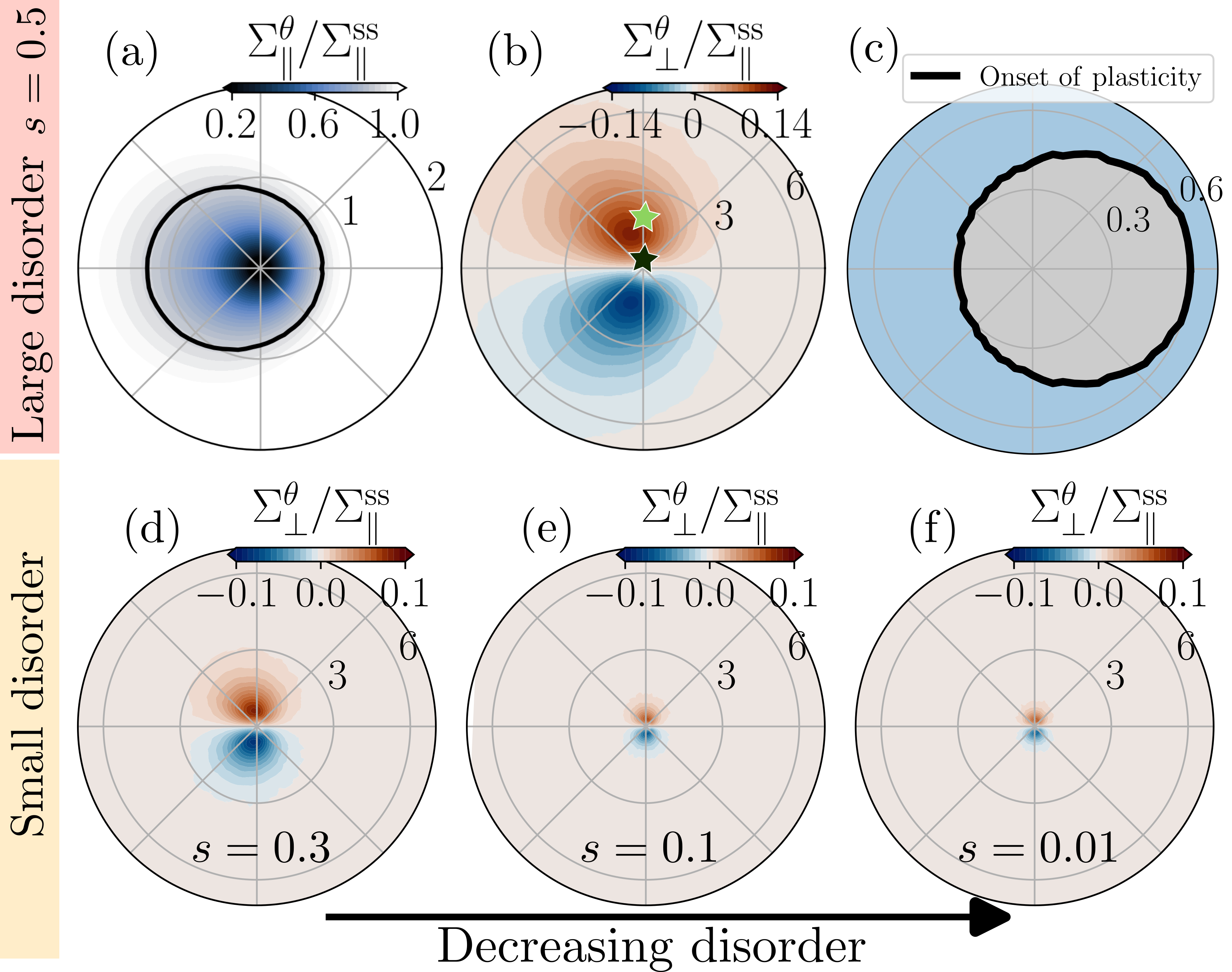}
\caption{Two-component Hébraud-Lequeux model. (a)--(c) Macroscopic behavior after rotation of the flow direction for large microscopic disorder ($s=0.5$): (a) rescaled shear stresses $\Sigma^\theta_\parallel/\Sigma^\mathrm{ss}_\parallel$  and (b) $\Sigma^\theta_\perp/\Sigma^\mathrm{ss}_\parallel$ and (c) Onset of plasticity $\Sigma^\mathrm{\theta}_{\mathrm{p}}/\Sigma^\mathrm{d}_\mathrm{y}$. (d)--(f) Same as (b), but for different disorder strengths: $s=0.3$ (d), $s=0.1$ (e) and $s=0.01$ (f).}
        \label{fig:2dhl_lognormal}
\end{figure}

To reproduce the experimental protocol with our model, we first apply a shear $\bm{E} = (\dot\gamma, 0)$ during 10 strain units, which is enough to reach steady state.
We then stop the shear by setting $\bm{E} = (0, 0)$ (keeping the strain fixed) to let the system relax to a rest state with $\Gamma=0$.
This state still has a finite macroscopic residual stress. 
We then unload the sample by applying a quasistatic reverse shear with $\bm{E} = (0^{-}, 0)$ to reach a relaxed state with vanishing macroscopic residual stress, $\bm{\Sigma} = 0$.
From this state, 
we apply a shear $\bm{E} = (\dot\gamma\cos\theta, -\dot\gamma\sin\theta)$, and track the tangential stress $\Sigma_\parallel = \Sigma_{xy}\cos\theta - \Sigma_{zy}\sin\theta$ and orthogonal stress $\Sigma_\perp = \Sigma_{xy}\sin\theta + \Sigma_{zy}\cos\theta$.
The shear rate $\dot\gamma$ is chosen to mimic the experiments: it ensures that the steady shear stress is 18\% higher than the dynamic yield stress $\Sigma^\mathrm{d}_\mathrm{y}$. The value of $\Sigma^\mathrm{d}_\mathrm{y}$ is determined by fitting the flow curve to a Herschel–Bulkley law $\Sigma^\mathrm{ss}_\parallel = \Sigma^\mathrm{d}_\mathrm{y} + k\dot\gamma^{0.5}$.

The macroscopic response to flow rotation is shown in Fig.~\ref{fig:2dhl_lognormal}a--d, in the case of a log-normal distribution with mean $0$ and standard deviation $s=0.5$ for the local yield stress distribution $\rho(\sigma_\mathrm{c})$, and a Gaussian distribution with standard deviation $1/3$ and zero mean for the resetting distribution $\rho_\mathrm{r}(\bm{\sigma})$. 
With these choices, the model captures the experimental phenomenology very well.
The post-rotation shear stresses $\Sigma^\theta_\parallel/\Sigma^\mathrm{ss}_\parallel$ (Fig.~\ref{fig:2dhl_lognormal}a) and  $\Sigma^\theta_\perp/\Sigma^\mathrm{ss}_\parallel$ (Fig.~\ref{fig:2dhl_lognormal}b) mirror the experimental results shown in Fig.~\ref{fig:contour_F12_F32_exp}.
The model predicts an anisotropic strain hardening (Fig.~\ref{fig:2dhl_lognormal}a) on strain scales comparable to the ones seen in experiments (Fig.~\ref{fig:contour_F12_F32_exp}a).
It also shows a transient behavior for $\Sigma^\theta_\perp$, with a two-lobe structure, $\Sigma^\theta_\perp>0$ for $0<\theta<\pi$ and $\Sigma^\theta_\perp<0$ for $-\pi<\theta<0$, here too in good agreement with the experiments (Fig.~\ref{fig:contour_F12_F32_exp}b). 
Finally, the transient regime extends to much larger strains for $\Sigma^\theta_\perp$, for which steady state is recovered after strains of order $\approx \numrange{5}{6}$, than for $\Sigma^\theta_\parallel$, for which steady state is reached after a strain of order 1.

The yield surface (Fig.~\ref{fig:2dhl_lognormal}c)
is also comparable to the one observed in the experiments (Fig.~\ref{fig:contour_F12_F32_exp}c).
The onset of plasticity $\Sigma^\theta_\mathrm{p}$ is roughly twice as large in the forward direction than it is in the backward direction.
The yield surface maintains a close to circular shape. 
This suggests that the effect of preshear on the anisotropy of the yield surface can be well described as pure kinematic hardening, that is, a mere translation of the center of the yield surface in the direction of applied flow~\cite{dimitriou2019canonical}.

We now turn more specifically to the role of the yield stress distribution. 
The duration of the transient for $\Sigma^\theta_\perp$ is strongly influenced by the width of the yield stress distribution $\rho(\sigma_\mathrm{c})$, as shown in Fig.~\ref{fig:2dhl_lognormal}d--f.
Indeed, decreasing the standard deviation $s$ of the log-normal law  to $s=0.3$ (Fig.~\ref{fig:2dhl_lognormal}d), $s=0.1$ (Fig.~\ref{fig:2dhl_lognormal}e), and $s=0.01$ (Fig.~\ref{fig:2dhl_lognormal}f), we observe that the strain needed for $\Sigma^\theta_\perp$ to reach its steady value sharply decreases.
By comparison, the transient for $\Sigma^\theta_\parallel$ also shortens, but in a much milder way, so that for small values of $s$, $\Sigma^\theta_\parallel$ and $\Sigma^\theta_\perp$ evolve over similar strain scales.

\begin{figure}
    \centering
        \includegraphics[width=\columnwidth]{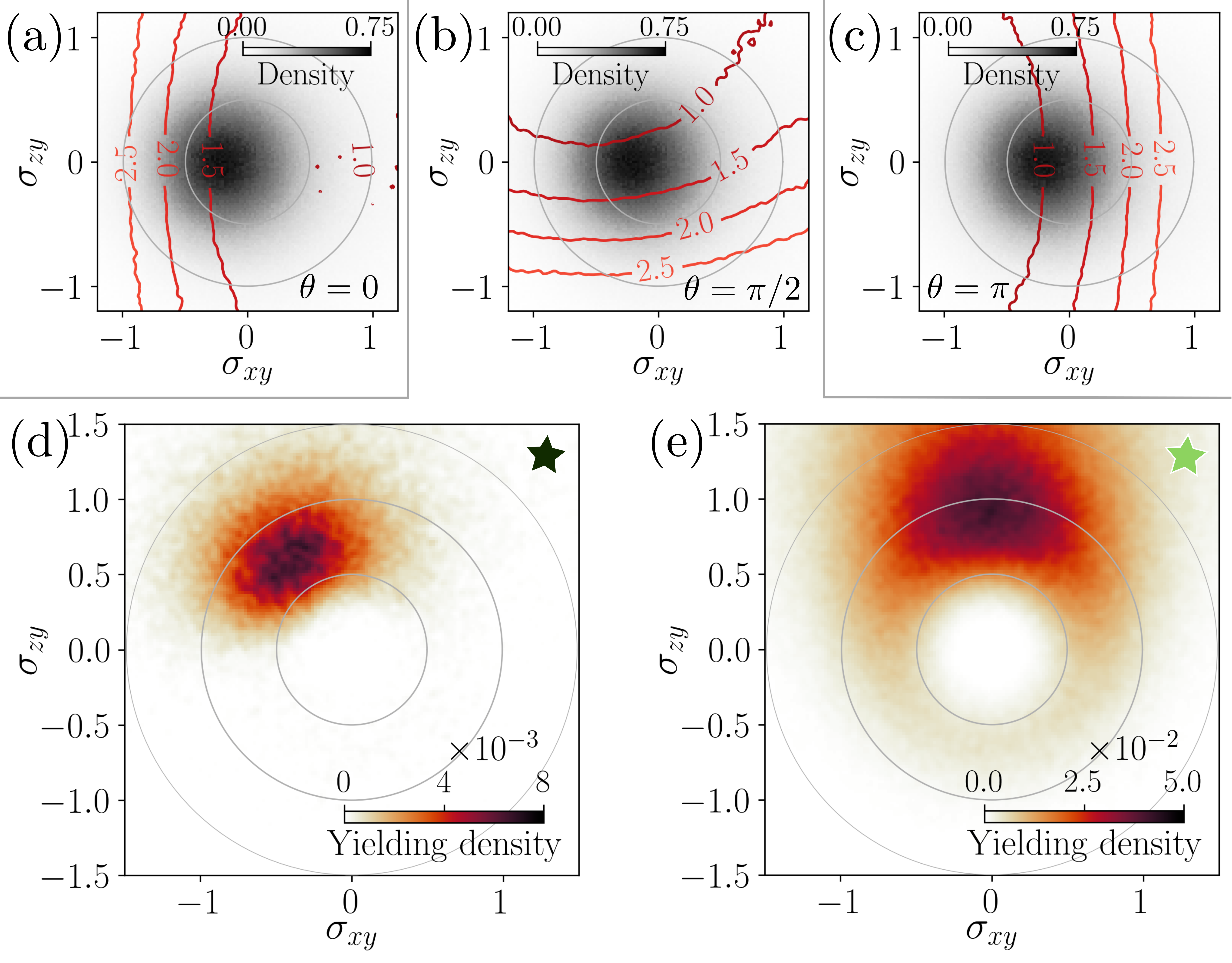}
   \caption{Origin of the $\Sigma_\perp$ behavior and kinematic hardening. (a) Stress distribution in the unloaded state (in grey) with iso-lines of distance to yield $x_\theta$ for $\theta=0$. (b) Same for $\theta=\pi/2$. (c) Same for $\theta=\pi$. (d) Density of sites susceptible to yield (with $|\bm{\sigma}| > \sigma_\mathrm{c}$) at a post-rotation strain $0.1$ for $\theta=\pi/2$. (e) Same, but at a post-rotation strain $2$. }
        \label{fig:2dhl_xmap}
\end{figure}

As in the experiments, the onset of plasticity corresponds to the onset of $\Sigma^\theta_\perp\neq0$ (see Supporting Information Fig.~S1b). The intriguing development of stress components $\Sigma^\theta_\perp$ perpendicular to the flow and the strain hardening phenomenon are thus intimately related, as we demonstrate below.  
The basic mechanism is the one we anticipated in the previous section.
During the pre-shear, a steady-state distribution $\mathcal{P}^\mathrm{ss}(\bm{\sigma}, \sigma_\mathrm{c})$ is established.
In the unloading phase, there is some plastic activity during the stress relaxation, but it is small enough to be ignored when trying to understand the overall qualitative behavior.
Hence the distribution $\mathcal{P}_0(\bm{\sigma}, \sigma_\mathrm{c})$ after unloading is essentially $\mathcal{P}^\mathrm{ss}$ translated by $-(\Sigma^\mathrm{ss}_\parallel, 0)$, 
so that $0 = \langle \bm{\sigma} \rangle = \int \mathrm{d}\bm{\sigma}\, \mathrm{d}\sigma_\mathrm{c}\, \mathcal{P}_0(\bm{\sigma}, \sigma_\mathrm{c}) \bm{\sigma}$.
Because of this translation to lower $\sigma_{xy}$ values, the sites with large positive $\sigma_{xy}$, which were typically close to yield in the forward direction during the preshear, are now further away from their plasticity threshold.
On the contrary, sites with initially small positive or negative $\sigma_{xy}$ are brought closer to their plasticity threshold in the backward direction.
When applying rotation of the flow direction, the parallel and orthogonal stresses are mixed.
To assess the consequences of the unloading on the subsequent shear at a given $\theta$, we compute the average distance to yield in direction $\theta$ for sites at a given $\bm{\sigma}$ as $\bar{x}_\theta(\bm{\sigma}) = \int \mathrm{d}\sigma_\mathrm{c} \mathcal{P}(\bm{\sigma}, \sigma_\mathrm{c}) x_\theta(\bm{\sigma}, \sigma_\mathrm{c})$ with $x_\theta(\bm{\sigma}, \sigma_\mathrm{c}) = \sqrt{\sigma_\mathrm{c}^2 - \sigma_\perp^2} - \sigma_\parallel$.
In Fig.~\ref{fig:2dhl_xmap}a--c, we show in grey scale the reduced stress distribution $\bar{\mathcal{P}_0}(\bm{\sigma}) = \int \mathrm{d}\sigma_\mathrm{c} \mathcal{P}_0(\bm{\sigma}, \sigma_\mathrm{c})$ showing the  asymmetry inherited from the steady state distribution $\mathcal{P}^\mathrm{ss}$.
We then overlay isolines of constant $\bar{x}_\theta$  (red lines)  for $\theta=0$ (Fig.~\ref{fig:2dhl_xmap}a), $\theta=\pi/2$ (Fig.~\ref{fig:2dhl_xmap}b) and $\theta=\pi$ (Fig.~\ref{fig:2dhl_xmap}c).
The effect of the unloading is clear: sites are typically much further to yield in the forward direction $\theta=0$ than in the backward direction $\theta=\pi$, whereas the $\theta=\pi/2$ is an in-between case. 
This explains the kinematic hardening observed in Fig.~\ref{fig:2dhl_lognormal}b--c.

Now focusing on the $\theta=\pi/2$ case, Fig.~\ref{fig:2dhl_xmap}b, among sites at a given $\sigma_{zy}$, sites at negative $\sigma_{xy}$ are much closer to yield that the ones at positive $\sigma_{xy}$.
Unsurprisingly, upon shear in the $\theta=\pi/2$ direction yielding occurs first in sites with $\sigma_\perp = \sigma_{xy} <0$, as shown in Fig.~\ref{fig:2dhl_xmap}d for a post-rotation strain of 0.1 (dark green star in Fig.~\ref{fig:2dhl_lognormal}b).
Depletion of sites with $\sigma_\perp<0$ leads to a positive macroscopic $\Sigma_\perp$.
This effect is however transient as the resetting of stress after a plastic event is isotropic, so that the initial asymmetry in $\bar{x}_\theta$ gradually disappears when yielding occurs for most sites.
This is visible in the yielding density shown in Fig.~\ref{fig:2dhl_xmap}e for a post-rotation strain of 2 (light green star in Fig.~\ref{fig:2dhl_lognormal}b).
Of course, the mirror situation occurs for $\theta=-\pi/2$, except that now $\sigma_\perp = -\sigma_{xy}$, so that the initial yielding of sites with $\sigma_{xy}<0$ yields to  $\Sigma_\perp<0$.

Hence, the 2-component elasto-plastic model shows that the fore-aft asymmetry of the local stress distributed built up during pre-shear, which was already argued to be at the origin of the scalar Bauschinger effect~\cite{patinet2020origin}, is also responsible for the finite orthogonal stress after change of shear direction.
This suggests that constitutive models designed to capture kinematic hardening may also be able to predict non-trivial orthogonal stresses after rotation of the flow direction.
We could confirm this with the model of Dimitriou \& McKinley~\cite{dimitriou2013describing,dimitriou2019canonical} (see Supporting Information for a brief description of the models and parameters we used), although the predicted behavior is not quite in quantitative agreement with our measurements (see Supporting Information Fig.~S2).

\section*{\bf Conclusion}

We have developed a novel shear experiment designed to study the anisotropy of the mechanical response of a soft glassy material as a function of its deformation history. Our apparatus can impose a simple shear in an arbitrary in-plane direction while simultaneously tracking the full tangential force vector. By systematically probing the response of the material at rest following a controlled preshear, we find that the material generically develops a non-trivial stress component orthogonal to the applied shear direction, in addition to the expected tangential component. We show that this behavior originates from an anisotropic internal stress distribution in the arrested state and is therefore intimately linked to the strain hardening observed and characterized for arbitrary shear directions.

Our mesoscopic modeling reveals that a key factor governing the transient orthogonal stress response in YSFs is the distribution of internal yield thresholds.
It shows that our rheology experiment, and more generally rheology involving complex deformation histories beyond simple shear or oscillatory shear, can inform us about the underlying disorder and its dynamical evolution.
Making 2D rheology a quantitative probe will require more systematic developments including spatialized elasto-plastic models equipped with accurate tensorial mesoscopic yield criteria.
Such criteria could be characterized by molecular dynamics or advanced mesoscale probing techniques~\cite{aimeUnifiedStateDiagram2023}.

Another example of complex fluids in the highly viscous regime that exhibit stress–flow misalignment are certain viscoelastic systems. In such materials, a finite stress-relaxation times can cause the response to depend on the deformation history, resulting in a transient misalignment between applied forces and the resulting deformations \cite{giesekus1982simple, ithaca1986theory}. This memory effect can give rise to rotated stress fields and transverse forces, as exemplified by the memory-induced Magnus effect \cite{cao2023memory}. 

In contrast, in our study, the structural disorder is unable to relax, persisting in an arrested state. This behavior is more reminiscent of plastic memory formation in solids, such as the (inverse) Taylor-Swift effect observed in crystalline materials, where the anisotropic internal structure leads to a persistent misalignment between stress and deformation under torsional loading \cite{taylor1934mechanism, swift1947elastic, molinari1997self}.

Thus, on a more fundamental level, our findings show that soft glassy materials may exhibit non-trivial plastic effects somewhat akin to the ones found in hard crystalline or amorphous materials exhibiting deformation induced anisotropies~\cite{sunFlowinducedElasticAnisotropy2016}.
This reinforces once more the status of soft glasses as athermal analogues to molecular glasses having the advantage of being easier to characterized experimentally at the microscale, via dynamic light scattering~\cite{aimeUnifiedStateDiagram2023} or tomography~\cite{schottMultiscaleStressDynamics2024}.

Our results also open the possibility for microscopically-informed constitutive models for engineered YSFs. 
We show that a strain hardening model is able to partially capture our observations~\cite{dimitriou2019canonical}, and we believe that our experimental setup, being a stringent test ground for such model, could prove extremely useful to develop refined models of this type.
An exciting perspective is that these improvement could enable the development of novel design strategies through deformation histories (similar to forging for hard and brittle solids), to create materials with programmable non-trivial mechanical responses.

\section{Appendix: methods}

\noindent\textit{Experiments}

The two-dimensional rheometer (2dR)\cite{blancRheologyDenseSuspensions2023} consists of two parallel plates mounted on orthogonally actuated translation stages (Newport MFA-CC, travel range 25~mm, resolution 100~nm) (Fig.~\ref{fig:CR_setup}). This configuration enables controlled relative motion between two PMMA plates in any planar direction, thereby allowing the imposition of simple shear with a velocity gradient normal to the plates. The resulting velocity gradient tensor is given by $\bm{L} = \dot\gamma \bm{e}_1 \bm{e}_2$, where $\bm{e}_1$, $\bm{e}_2$, and $\bm{e}_3$ denote the flow, gradient, and vorticity directions, respectively. The shear rate $\dot\gamma$ is defined from the velocity of the upper plate relative to the lower one: $\bm{v} = \dot\gamma d \bm{e}_1$, with $d$ the gap between the plates. A multi-axis force sensor (AMTI HE6x6-1, maximum load $\SI{2.2}{\newton}$, resolution \SI{2}{\milli\newton}) mounted beneath the lower plate measures the shear force transmitted by the material, capturing both its magnitude and orientation: $\bm{F} = F_{\parallel} \bm{e}_1 + F_{\perp} \bm{e}_3$ (Fig.~\ref{fig:CR_setup}b–c).

In the experiments, the gap is set to $d=\SI{1}{\milli\meter}$, the bottom plate diameter is $\SI{160}{\milli\meter}$, and the upper plate diameter is $\SI{80}{\milli\meter}$. When the gap is completely filled with a material, with these parameters, the 2dR achieves a strain resolution of approximately 0.001 and a stress resolution of about \SI{1}{Pa}, with typical stress magnitudes near \SI{100}{Pa}. Initially designed to study the tensorial viscosity of dense non-Brownian suspensions under rotation of the flow direction, the 2dR has been instrumental in revealing the distinct contributions of contact and hydrodynamic forces to the overall rheology~\cite{blancRheologyDenseSuspensions2023}.

In the present study, we employ the 2dR to investigate the elastoplastic behavior of a yield stress fluid, a Carbopol gel. 
Carbopol 980 was obtained from Lubrizol in powder form, and gel was prepared at 2~wt\% concentration.
The sample was prepared by slowly dispersing the powder in Milli-Q ultrapure water ($\SI{18.2}{\mega \ohm \centi \meter}$) under mechanical stirring using a three-blade marine impeller (diameter \SI{4}{\centi\meter}) in a \SI{250}{\milli \liter} beaker. Moderate agitation (\SI{800}{rpm} - IKA Eurostar 200 Control) was initially applied; the material was then stirred at \SI{400}{rpm} for two hours to allow the Carbopol to hydrate. The resulting pH was close to 3.
The sample was further neutralized to pH~7 using a 1M NaOH commercial solution (Fisher Chemical), leading to the swelling of the microgel particles and the formation of a gel, and was vigorously mixed for 10~minutes at \SI{1500}{rpm} using a planetary centrifugal mixer (Thinky ARE-250). This type of mixer combines revolution and rotary movement of the container, which not only mixes the sample but also helps remove bubbles entrapped within the fluid. All samples were allowed to rest overnight at room temperature before further characterization to allow the pH to stabilize and reach equilibrium. 

\begin{figure}
    \centering
    \includegraphics[width=\linewidth]{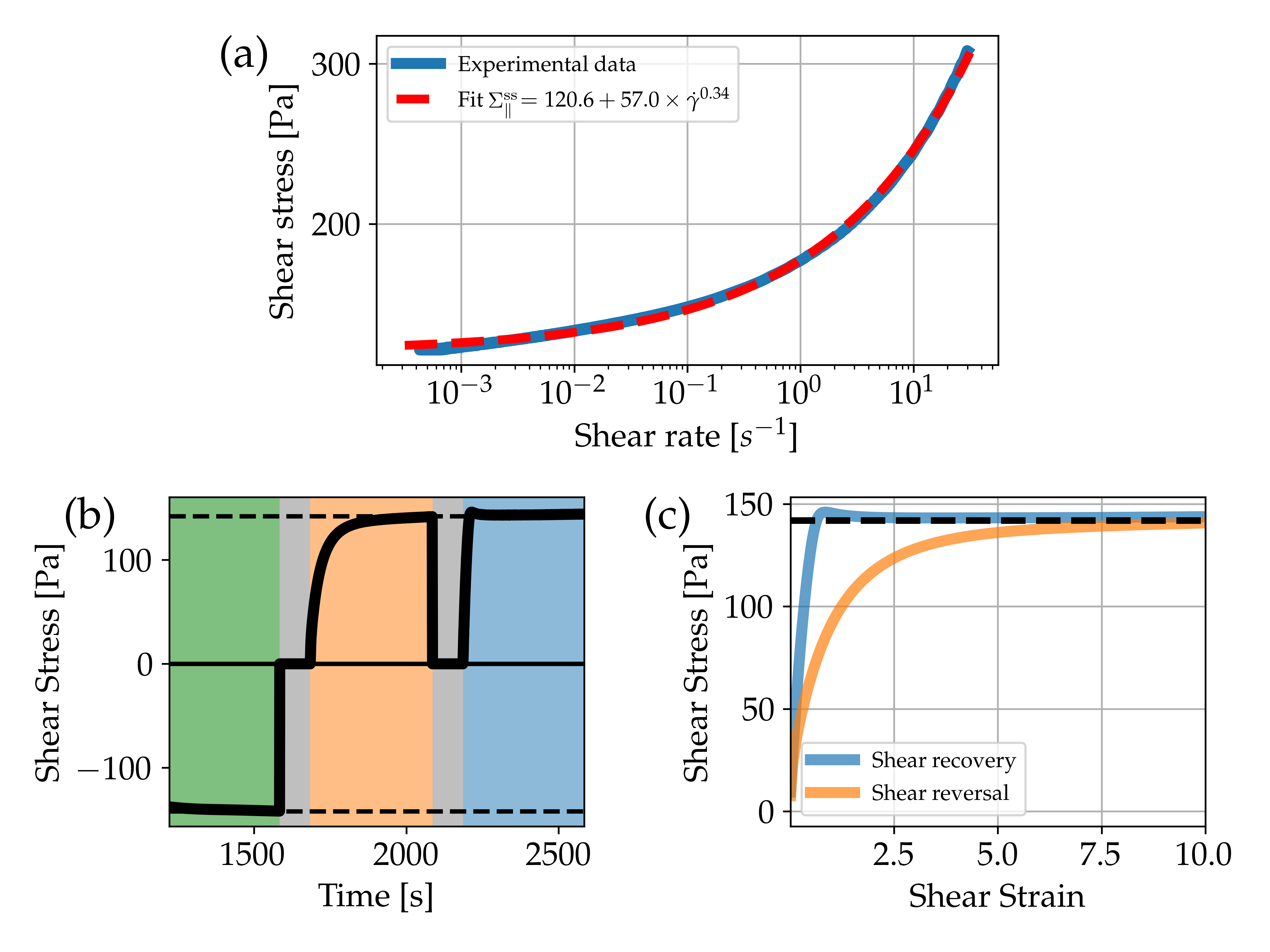}
    \caption{Rheological behavior of the Carbopol gel.  
    (a) Flow curve obtained by progressively decreasing the shear rate from 50~$\mathrm{s^{-1}}$ to $10^{-3}~\mathrm{s^{-1}}$ over \SI{570}{\second}.  
    Blue line: experimental data for the flow curve. Red dashed line: fit to a Herschel-Bulkley model $\Sigma^\mathrm{ss}_\parallel = \Sigma_y^d + k \dot\gamma^n$.  
    (b) Elastoplastic response of the gel. A preshear is applied at a constant shear rate of $-\SI{0.03}{\per\second}$ (green window), followed by a rest period of \SI{30}{\second} (grey window). Shear is then reversed at $+\SI{0.03}{\per\second}$ (orange window) to characterize the response to shear reversal. This is followed by another rest period at zero stress (grey window) and by another shear in the same direction at the same shear rate (blue window) to characterize the response to shear recovery.   
    (c) Same data as in (b), showing shear stress versus shear strain for shear reversal (orange curve) and shear recovery (blue curve) experiments. The black dashed line in (b) and (c) indicates the stress level $1.18\Sigma_y^d$.}
    \label{fig:rheo_macro}
\end{figure}

The rheology of the sample was characterized at a constant temperature of 25°C using a ThermoFischer Mars II rheometer equipped with sandblasted parallel plate geometry (\SI{60}{\milli \meter}, \SI{1}{\milli\meter} gap). The material was first presheared for \SI{30}{\second} at \SI{50}{\per\second}, and a flow curve was obtained by applying a decreasing shear rate ramp from 50~$\mathrm{s^{-1}}$ to $10^{-3}~\mathrm{s^{-1}}$ in \SI{570}{\second} (Fig.~\ref{fig:rheo_macro}a). The studied Carbopol gel is characterized by a Herschel-Bulkley behavior $\Sigma=\Sigma_y^d+k\dot\gamma^n$, with $\Sigma_y^d=\SI{120.6}{\pascal}$, $k=\SI{57.0}{\pascal\,\second^{n}}$ and $n=\SI{0.34}{}$. The elastoplastic response of the gel was characterized by imposing a steady shear rate of \SI{0.03}{\per \second} after a preshear in the two opposite directions allowed by the rheometer. Its elastoplastic behavior shows strong Bauschinger effect (Fig.~\ref{fig:rheo_macro}b--c).

Prior to experiments with the 2D rheometer setup, the gel is carefully deposited at the center of the bottom plate.
The upper plate is then lowered to set the gap to \SI{1}{\milli\meter}, inducing a squeeze flow that spreads the gel into a cylindrical shape.
The resulting contact area is directly observable through the transparent upper plate.
During this loading phase, both vertical and horizontal forces increase due to the induced deformation.
To relax these residual stresses before performing the experiments, we apply a mechanical preconditioning protocol consisting of shear in multiple directions with gradually increasing strain amplitude. 
   
To suppress wall slip in the 2dR, the PMMA plates were coated with a thin layer of polyethylenimine (PEI)~\cite{christel2012stick}. To suppress water loss through evaporation of the Carbopol gel during experimentation, a thin layer of oil was carefully applied around the sample.

A typical experimental sequence begins with a preshear phase to reach a steady-state, followed by stress relaxation and by a controlled rotation of the flow direction. This rotation consists in changing the flow direction $\bm{e}_1$ and the vorticity direction $\bm{e}_3$ by an angle $\theta \in [-\pi, \pi]$ around the gradient direction $\bm{e}_2$ (see thick blue arrow in Fig.~\ref{fig:CR_setup}c). We then monitor the force components $F_\parallel(\gamma,\theta)$ and $F_\perp(\gamma,\theta)$ as functions of the strain $\gamma$ accumulated after the rotation, in order to probe the rheological response as a function of shear history.

To systematically explore the elastoplastic and strain-hardening behavior of this yield stress fluid~\cite{deboeufMechanismStrainHardening2022}, we follow a reproducible protocol. 
A preliminary shear is applied at a constant rate of $3 \times 10^{-2}$~s$^{-1}$ up to a total strain of approximately 20, effectively erasing prior shear history. This is followed by a reverse shear at the same rate to fully relax the macroscopic stress. After a subsequent rest period, shear is re-applied in rotated directions defined by angles $\theta$, with each test involving a total strain of 7. By incrementing $\theta$ in steps of $3^\circ$, we construct a detailed two-dimensional map of the material’s directional response (Fig.~\ref{fig:contour_F12_F32_exp}).

\ \\ \noindent\textit{Model}

To solve the coupled Langevin equations Eq.~\ref{eq:langevin} and Eq.~\ref{eq:langevin_yield} numerically we discretize the time using the It\^o scheme~\cite{riskenFokkerPlanckEquationMethods1996}.
In the limit of a large number of random walkers, our minimal model reproduces the dynamics of a two-stress-component generalization of the disordered Hébraud-Lequeux model~\cite{hebraud1998mode,agoritsasRelevanceDisorderAthermal2015}:
The evolution of the probability distribution function 
$\mathcal{P}(\bm{\sigma}, \sigma_\mathrm{c}, t)$ of having local stress components $(\sigma_{xy}, \sigma_{zy}) \equiv \bm{\sigma}$ and a local yield stress $\sigma_\mathrm{c}$ at time $t$ evolves under an imposed flow $\bm{E} = (E_{xy}, E_{zy}) = (\dot\gamma \cos\theta, -\dot\gamma \sin\theta)$ as follows
\begin{multline}\label{eq:2dHL_model}
    \partial_t \mathcal{P}(\bm{\sigma}, \sigma_\mathrm{c}, t) = -G \bm{E}.\nabla_{\bm{\sigma}} \mathcal{P} + D(t)\nabla^2_{\bm{\sigma}}\mathcal{P} \\ 
    - \tau^{-1} \Theta (\lvert \bm{\sigma} \rvert - \sigma^c)\mathcal{P} + \Gamma (t) \rho_\mathrm{r}(\bm{\sigma}) \rho(\sigma_\mathrm{c})\;.
\end{multline}
The first two r.h.s. terms respectively describe the elastic loading under shear and the mechanical noise on the local stress generated by plastic events in the system.
Note that there is no convective derivative term as it does not couple $\sigma_{xy}$ and $\sigma_{zy}$ under the imposed flow we consider.
The last two terms capture the yielding process: 
 after a plastic event the stress is drawn from a distribution $\rho_\mathrm{r}(\bm{\sigma})$, while a new local yield stress is drawn from a distribution $\rho(\sigma_\mathrm{c})$. 
Here, we assume a local von Mises yield criterion, that is, yield occurs for $\lvert \bm{\sigma} \rvert = \sqrt{\sigma_{xy}^2 + \sigma_{zy}^2} \geq \sigma^c$.
Furthermore, yield is not instantaneous, it occurs stochastically with a rate $\tau^{-1}$.
Probability conservation imposes that 
\begin{equation}
    \Gamma(t) = \int_0^{\infty} d \sigma_\mathrm{c} \int_{\lvert \bm{\sigma} \rvert > \sigma^c} d^2 \bm{\sigma} \mathcal{P}(\bm{\sigma}, \sigma_\mathrm{c}, t) \, .
\end{equation}
Finally, the mechanical noise is proportional to the proportion of plastic sites
\begin{equation}
    D(t) = \alpha \Gamma(t)\, ,
\end{equation}
with $\alpha$ a constant which controls the steady-state rheological properties. For $\alpha < 1/4$, the model describes a yield stress fluid, while for $\alpha \geq 1/4$ the yield stress disappears~\cite{olivierGeneralizationHebraudLequeux2013}.
Here, we set $\alpha$ to the value $0.15$.



\section*{Acknowledgements}
We thank Julien Olivier for helpful discussions on the generalized H\'ebraud-Lequeux model.
This work was supported by the French ANR, grant number ANR-24-CE06-1565 (HistoRYS project).




\begin{thebibliography}{78}%
\makeatletter
\providecommand \@ifxundefined [1]{%
 \@ifx{#1\undefined}
}%
\providecommand \@ifnum [1]{%
 \ifnum #1\expandafter \@firstoftwo
 \else \expandafter \@secondoftwo
 \fi
}%
\providecommand \@ifx [1]{%
 \ifx #1\expandafter \@firstoftwo
 \else \expandafter \@secondoftwo
 \fi
}%
\providecommand \natexlab [1]{#1}%
\providecommand \enquote  [1]{``#1''}%
\providecommand \bibnamefont  [1]{#1}%
\providecommand \bibfnamefont [1]{#1}%
\providecommand \citenamefont [1]{#1}%
\providecommand \href@noop [0]{\@secondoftwo}%
\providecommand \href [0]{\begingroup \@sanitize@url \@href}%
\providecommand \@href[1]{\@@startlink{#1}\@@href}%
\providecommand \@@href[1]{\endgroup#1\@@endlink}%
\providecommand \@sanitize@url [0]{\catcode `\\12\catcode `\$12\catcode
  `\&12\catcode `\#12\catcode `\^12\catcode `\_12\catcode `\%12\relax}%
\providecommand \@@startlink[1]{}%
\providecommand \@@endlink[0]{}%
\providecommand \url  [0]{\begingroup\@sanitize@url \@url }%
\providecommand \@url [1]{\endgroup\@href {#1}{\urlprefix }}%
\providecommand \urlprefix  [0]{URL }%
\providecommand \Eprint [0]{\href }%
\providecommand \doibase [0]{https://doi.org/}%
\providecommand \selectlanguage [0]{\@gobble}%
\providecommand \bibinfo  [0]{\@secondoftwo}%
\providecommand \bibfield  [0]{\@secondoftwo}%
\providecommand \translation [1]{[#1]}%
\providecommand \BibitemOpen [0]{}%
\providecommand \bibitemStop [0]{}%
\providecommand \bibitemNoStop [0]{.\EOS\space}%
\providecommand \EOS [0]{\spacefactor3000\relax}%
\providecommand \BibitemShut  [1]{\csname bibitem#1\endcsname}%
\let\auto@bib@innerbib\@empty
\bibitem [{\citenamefont {Coussot}(2014)}]{coussot2014yield}%
  \BibitemOpen
  \bibfield  {author} {\bibinfo {author} {\bibfnamefont {P.}~\bibnamefont
  {Coussot}},\ }\bibfield  {title} {\bibinfo {title} {Yield stress fluid flows:
  A review of experimental data},\ }\href@noop {} {\bibfield  {journal}
  {\bibinfo  {journal} {Journal of Non-Newtonian Fluid Mechanics}\ }\textbf
  {\bibinfo {volume} {211}},\ \bibinfo {pages} {31} (\bibinfo {year}
  {2014})}\BibitemShut {NoStop}%
\bibitem [{\citenamefont {Bonn}\ \emph {et~al.}(2017)\citenamefont {Bonn},
  \citenamefont {Denn}, \citenamefont {Berthier}, \citenamefont {Divoux},\ and\
  \citenamefont {Manneville}}]{bonnYieldStressMaterials2017}%
  \BibitemOpen
  \bibfield  {author} {\bibinfo {author} {\bibfnamefont {D.}~\bibnamefont
  {Bonn}}, \bibinfo {author} {\bibfnamefont {M.~M.}\ \bibnamefont {Denn}},
  \bibinfo {author} {\bibfnamefont {L.}~\bibnamefont {Berthier}}, \bibinfo
  {author} {\bibfnamefont {T.}~\bibnamefont {Divoux}},\ and\ \bibinfo {author}
  {\bibfnamefont {S.}~\bibnamefont {Manneville}},\ }\bibfield  {title}
  {\bibinfo {title} {Yield stress materials in soft condensed matter},\ }\href
  {https://doi.org/10.1103/RevModPhys.89.035005} {\bibfield  {journal}
  {\bibinfo  {journal} {Reviews of Modern Physics}\ }\textbf {\bibinfo {volume}
  {89}},\ \bibinfo {pages} {035005} (\bibinfo {year} {2017})},\ \Eprint
  {https://arxiv.org/abs/1502.05281} {1502.05281} \BibitemShut {NoStop}%
\bibitem [{\citenamefont {Balmforth}\ \emph {et~al.}(2014)\citenamefont
  {Balmforth}, \citenamefont {Frigaard},\ and\ \citenamefont
  {Ovarlez}}]{balmforth2014yielding}%
  \BibitemOpen
  \bibfield  {author} {\bibinfo {author} {\bibfnamefont {N.~J.}\ \bibnamefont
  {Balmforth}}, \bibinfo {author} {\bibfnamefont {I.~A.}\ \bibnamefont
  {Frigaard}},\ and\ \bibinfo {author} {\bibfnamefont {G.}~\bibnamefont
  {Ovarlez}},\ }\bibfield  {title} {\bibinfo {title} {Yielding to stress:
  recent developments in viscoplastic fluid mechanics},\ }\href
  {https://doi.org/10.1146/annurev-fluid-010313-141424} {\bibfield  {journal}
  {\bibinfo  {journal} {Annual Review of Fluid Mechanics}\ }\textbf {\bibinfo
  {volume} {46}},\ \bibinfo {pages} {121} (\bibinfo {year} {2014})},\ \Eprint
  {https://arxiv.org/abs/https://personal.math.ubc.ca/~njb/Research/annrev.pdf}
  {https://personal.math.ubc.ca/~njb/Research/annrev.pdf} \BibitemShut
  {NoStop}%
\bibitem [{\citenamefont {Hunter}\ and\ \citenamefont
  {Weeks}(2012)}]{Hunter_2012}%
  \BibitemOpen
  \bibfield  {author} {\bibinfo {author} {\bibfnamefont {G.~L.}\ \bibnamefont
  {Hunter}}\ and\ \bibinfo {author} {\bibfnamefont {E.~R.}\ \bibnamefont
  {Weeks}},\ }\bibfield  {title} {\bibinfo {title} {The physics of the
  colloidal glass transition},\ }\href
  {https://doi.org/10.1088/0034-4885/75/6/066501} {\bibfield  {journal}
  {\bibinfo  {journal} {Reports on Progress in Physics}\ }\textbf {\bibinfo
  {volume} {75}},\ \bibinfo {pages} {066501} (\bibinfo {year}
  {2012})}\BibitemShut {NoStop}%
\bibitem [{\citenamefont {Nicolas}\ \emph {et~al.}(2018)\citenamefont
  {Nicolas}, \citenamefont {Ferrero}, \citenamefont {Martens},\ and\
  \citenamefont {Barrat}}]{nicolas2018deformation}%
  \BibitemOpen
  \bibfield  {author} {\bibinfo {author} {\bibfnamefont {A.}~\bibnamefont
  {Nicolas}}, \bibinfo {author} {\bibfnamefont {E.~E.}\ \bibnamefont
  {Ferrero}}, \bibinfo {author} {\bibfnamefont {K.}~\bibnamefont {Martens}},\
  and\ \bibinfo {author} {\bibfnamefont {J.~L.}\ \bibnamefont {Barrat}},\
  }\bibfield  {title} {\bibinfo {title} {Deformation and flow of amorphous
  solids: Insights from elastoplastic models},\ }\href@noop {} {\bibfield
  {journal} {\bibinfo  {journal} {Reviews of Modern Physics}\ }\textbf
  {\bibinfo {volume} {90}},\ \bibinfo {pages} {045006} (\bibinfo {year}
  {2018})}\BibitemShut {NoStop}%
\bibitem [{\citenamefont {Divoux}\ \emph {et~al.}(2024)\citenamefont {Divoux},
  \citenamefont {Agoritsas}, \citenamefont {Aime}, \citenamefont {Barentin},
  \citenamefont {Barrat}, \citenamefont {Benzi}, \citenamefont {Berthier},
  \citenamefont {Bi}, \citenamefont {Biroli}, \citenamefont {Bonn} \emph
  {et~al.}}]{divoux2024ductile}%
  \BibitemOpen
  \bibfield  {author} {\bibinfo {author} {\bibfnamefont {T.}~\bibnamefont
  {Divoux}}, \bibinfo {author} {\bibfnamefont {E.}~\bibnamefont {Agoritsas}},
  \bibinfo {author} {\bibfnamefont {S.}~\bibnamefont {Aime}}, \bibinfo {author}
  {\bibfnamefont {C.}~\bibnamefont {Barentin}}, \bibinfo {author}
  {\bibfnamefont {J.-L.}\ \bibnamefont {Barrat}}, \bibinfo {author}
  {\bibfnamefont {R.}~\bibnamefont {Benzi}}, \bibinfo {author} {\bibfnamefont
  {L.}~\bibnamefont {Berthier}}, \bibinfo {author} {\bibfnamefont
  {D.}~\bibnamefont {Bi}}, \bibinfo {author} {\bibfnamefont {G.}~\bibnamefont
  {Biroli}}, \bibinfo {author} {\bibfnamefont {D.}~\bibnamefont {Bonn}}, \emph
  {et~al.},\ }\bibfield  {title} {\bibinfo {title} {Ductile-to-brittle
  transition and yielding in soft amorphous materials: perspectives and open
  questions},\ }\href@noop {} {\bibfield  {journal} {\bibinfo  {journal} {Soft
  Matter}\ }\textbf {\bibinfo {volume} {20}},\ \bibinfo {pages} {6868}
  (\bibinfo {year} {2024})}\BibitemShut {NoStop}%
\bibitem [{\citenamefont {Saramito}(2007)}]{saramito2007new}%
  \BibitemOpen
  \bibfield  {author} {\bibinfo {author} {\bibfnamefont {P.}~\bibnamefont
  {Saramito}},\ }\bibfield  {title} {\bibinfo {title} {A new constitutive
  equation for elastoviscoplastic fluid flows},\ }\href@noop {} {\bibfield
  {journal} {\bibinfo  {journal} {J. Non-Newt. Fluid Mech.}\ }\textbf {\bibinfo
  {volume} {145}},\ \bibinfo {pages} {1} (\bibinfo {year} {2007})}\BibitemShut
  {NoStop}%
\bibitem [{\citenamefont
  {Saramito}(2009)}]{saramitoNewElastoviscoplasticModel2009}%
  \BibitemOpen
  \bibfield  {author} {\bibinfo {author} {\bibfnamefont {P.}~\bibnamefont
  {Saramito}},\ }\bibfield  {title} {\bibinfo {title} {A new elastoviscoplastic
  model based on the {{Herschel}}--{{Bulkley}} viscoplastic model},\ }\href
  {https://doi.org/10.1016/j.jnnfm.2008.12.001} {\bibfield  {journal} {\bibinfo
   {journal} {Journal of Non-Newtonian Fluid Mechanics}\ }\bibinfo {series}
  {Visco-Plastic Fluids: {{From}} Theory to Application},\ \textbf {\bibinfo
  {volume} {158}},\ \bibinfo {pages} {154} (\bibinfo {year}
  {2009})}\BibitemShut {NoStop}%
\bibitem [{\citenamefont {Ovarlez}\ \emph {et~al.}(2010)\citenamefont
  {Ovarlez}, \citenamefont {Barral},\ and\ \citenamefont
  {Coussot}}]{ovarlez2010three}%
  \BibitemOpen
  \bibfield  {author} {\bibinfo {author} {\bibfnamefont {G.}~\bibnamefont
  {Ovarlez}}, \bibinfo {author} {\bibfnamefont {Q.}~\bibnamefont {Barral}},\
  and\ \bibinfo {author} {\bibfnamefont {P.}~\bibnamefont {Coussot}},\
  }\bibfield  {title} {\bibinfo {title} {Three-dimensional jamming and flows of
  soft glassy materials},\ }\href {https://doi.org/10.1038/nmat2615} {\bibfield
   {journal} {\bibinfo  {journal} {Nature Materials}\ }\textbf {\bibinfo
  {volume} {9}},\ \bibinfo {pages} {115} (\bibinfo {year} {2010})},\ \Eprint
  {https://arxiv.org/abs/0910.1821} {0910.1821} \BibitemShut {NoStop}%
\bibitem [{\citenamefont {Kamani}\ \emph {et~al.}(2021)\citenamefont {Kamani},
  \citenamefont {Donley},\ and\ \citenamefont
  {Rogers}}]{kamani2021unification}%
  \BibitemOpen
  \bibfield  {author} {\bibinfo {author} {\bibfnamefont {K.}~\bibnamefont
  {Kamani}}, \bibinfo {author} {\bibfnamefont {G.~J.}\ \bibnamefont {Donley}},\
  and\ \bibinfo {author} {\bibfnamefont {S.~A.}\ \bibnamefont {Rogers}},\
  }\bibfield  {title} {\bibinfo {title} {Unification of the rheological physics
  of yield stress fluids},\ }\href@noop {} {\bibfield  {journal} {\bibinfo
  {journal} {Physical Review Letters}\ }\textbf {\bibinfo {volume} {126}},\
  \bibinfo {pages} {218002} (\bibinfo {year} {2021})}\BibitemShut {NoStop}%
\bibitem [{\citenamefont {Fraggedakis}\ \emph
  {et~al.}(2016{\natexlab{a}})\citenamefont {Fraggedakis}, \citenamefont
  {Dimakopoulos},\ and\ \citenamefont
  {Tsamopoulos}}]{fraggedakis2016yielding1}%
  \BibitemOpen
  \bibfield  {author} {\bibinfo {author} {\bibfnamefont {D.}~\bibnamefont
  {Fraggedakis}}, \bibinfo {author} {\bibfnamefont {Y.}~\bibnamefont
  {Dimakopoulos}},\ and\ \bibinfo {author} {\bibfnamefont {J.}~\bibnamefont
  {Tsamopoulos}},\ }\bibfield  {title} {\bibinfo {title} {Yielding the
  yield-stress analysis: a study focused on the effects of elasticity on the
  settling of a single spherical particle in simple yield-stress fluids},\
  }\href@noop {} {\bibfield  {journal} {\bibinfo  {journal} {Soft matter}\
  }\textbf {\bibinfo {volume} {12}},\ \bibinfo {pages} {5378} (\bibinfo {year}
  {2016}{\natexlab{a}})}\BibitemShut {NoStop}%
\bibitem [{\citenamefont {Fraggedakis}\ \emph
  {et~al.}(2016{\natexlab{b}})\citenamefont {Fraggedakis}, \citenamefont
  {Dimakopoulos},\ and\ \citenamefont
  {Tsamopoulos}}]{fraggedakis2016yielding2}%
  \BibitemOpen
  \bibfield  {author} {\bibinfo {author} {\bibfnamefont {D.}~\bibnamefont
  {Fraggedakis}}, \bibinfo {author} {\bibfnamefont {Y.}~\bibnamefont
  {Dimakopoulos}},\ and\ \bibinfo {author} {\bibfnamefont {J.}~\bibnamefont
  {Tsamopoulos}},\ }\bibfield  {title} {\bibinfo {title} {Yielding the yield
  stress analysis: A thorough comparison of recently proposed
  elasto-visco-plastic (evp) fluid models},\ }\href@noop {} {\bibfield
  {journal} {\bibinfo  {journal} {Journal of Non-Newtonian Fluid Mechanics}\
  }\textbf {\bibinfo {volume} {236}},\ \bibinfo {pages} {104} (\bibinfo {year}
  {2016}{\natexlab{b}})}\BibitemShut {NoStop}%
\bibitem [{\citenamefont {Cheddadi}\ \emph {et~al.}(2011)\citenamefont
  {Cheddadi}, \citenamefont {Saramito}, \citenamefont {Dollet}, \citenamefont
  {Raufaste},\ and\ \citenamefont
  {Graner}}]{cheddadiUnderstandingPredictingViscous2011}%
  \BibitemOpen
  \bibfield  {author} {\bibinfo {author} {\bibfnamefont {I.}~\bibnamefont
  {Cheddadi}}, \bibinfo {author} {\bibfnamefont {P.}~\bibnamefont {Saramito}},
  \bibinfo {author} {\bibfnamefont {B.}~\bibnamefont {Dollet}}, \bibinfo
  {author} {\bibfnamefont {C.}~\bibnamefont {Raufaste}},\ and\ \bibinfo
  {author} {\bibfnamefont {F.}~\bibnamefont {Graner}},\ }\bibfield  {title}
  {\bibinfo {title} {Understanding and predicting viscous, elastic, plastic
  flows},\ }\href {https://doi.org/10.1140/epje/i2011-11001-4} {\bibfield
  {journal} {\bibinfo  {journal} {The European Physical Journal E}\ }\textbf
  {\bibinfo {volume} {34}},\ \bibinfo {pages} {1} (\bibinfo {year}
  {2011})}\BibitemShut {NoStop}%
\bibitem [{\citenamefont {Kamani}\ and\ \citenamefont
  {Rogers}(2024)}]{kamani2024brittle}%
  \BibitemOpen
  \bibfield  {author} {\bibinfo {author} {\bibfnamefont {K.~M.}\ \bibnamefont
  {Kamani}}\ and\ \bibinfo {author} {\bibfnamefont {S.~A.}\ \bibnamefont
  {Rogers}},\ }\bibfield  {title} {\bibinfo {title} {Brittle and ductile
  yielding in soft materials},\ }\href@noop {} {\bibfield  {journal} {\bibinfo
  {journal} {Proceedings of the National Academy of Sciences}\ }\textbf
  {\bibinfo {volume} {121}},\ \bibinfo {pages} {e2401409121} (\bibinfo {year}
  {2024})}\BibitemShut {NoStop}%
\bibitem [{\citenamefont {Agrawal}\ \emph {et~al.}(2025)\citenamefont
  {Agrawal}, \citenamefont {Garc{\'\i}a-Tu{\~n}{\'o}n}, \citenamefont {Poole},\
  and\ \citenamefont {Fonte}}]{agrawal2025features}%
  \BibitemOpen
  \bibfield  {author} {\bibinfo {author} {\bibfnamefont {R.}~\bibnamefont
  {Agrawal}}, \bibinfo {author} {\bibfnamefont {E.}~\bibnamefont
  {Garc{\'\i}a-Tu{\~n}{\'o}n}}, \bibinfo {author} {\bibfnamefont {R.~J.}\
  \bibnamefont {Poole}},\ and\ \bibinfo {author} {\bibfnamefont {C.~P.}\
  \bibnamefont {Fonte}},\ }\bibfield  {title} {\bibinfo {title} {Features and
  limitations of recent elastoviscoplastic constitutive models under large
  amplitude oscillatory shear (laos)},\ }\href@noop {} {\bibfield  {journal}
  {\bibinfo  {journal} {Journal of Non-Newtonian Fluid Mechanics}\ }\textbf
  {\bibinfo {volume} {338}},\ \bibinfo {pages} {105407} (\bibinfo {year}
  {2025})}\BibitemShut {NoStop}%
\bibitem [{\citenamefont {Brader}\ \emph {et~al.}(2008)\citenamefont {Brader},
  \citenamefont {Cates},\ and\ \citenamefont
  {Fuchs}}]{braderFirstPrinciplesConstitutiveEquation2008}%
  \BibitemOpen
  \bibfield  {author} {\bibinfo {author} {\bibfnamefont {J.~M.}\ \bibnamefont
  {Brader}}, \bibinfo {author} {\bibfnamefont {M.~E.}\ \bibnamefont {Cates}},\
  and\ \bibinfo {author} {\bibfnamefont {M.}~\bibnamefont {Fuchs}},\ }\bibfield
   {title} {\bibinfo {title} {First-{{Principles Constitutive Equation}} for
  {{Suspension Rheology}}},\ }\href
  {https://doi.org/10.1103/PhysRevLett.101.138301} {\bibfield  {journal}
  {\bibinfo  {journal} {Physical Review Letters}\ }\textbf {\bibinfo {volume}
  {101}},\ \bibinfo {pages} {138301} (\bibinfo {year} {2008})}\BibitemShut
  {NoStop}%
\bibitem [{\citenamefont {Brader}\ \emph {et~al.}(2009)\citenamefont {Brader},
  \citenamefont {Voigtmann}, \citenamefont {Fuchs}, \citenamefont {Larson},\
  and\ \citenamefont {Cates}}]{braderGlassRheologyModecoupling2009}%
  \BibitemOpen
  \bibfield  {author} {\bibinfo {author} {\bibfnamefont {J.~M.}\ \bibnamefont
  {Brader}}, \bibinfo {author} {\bibfnamefont {T.}~\bibnamefont {Voigtmann}},
  \bibinfo {author} {\bibfnamefont {M.}~\bibnamefont {Fuchs}}, \bibinfo
  {author} {\bibfnamefont {R.~G.}\ \bibnamefont {Larson}},\ and\ \bibinfo
  {author} {\bibfnamefont {M.~E.}\ \bibnamefont {Cates}},\ }\bibfield  {title}
  {\bibinfo {title} {Glass rheology: {{From}} mode-coupling theory to a
  dynamical yield criterion},\ }\href {https://doi.org/10.1073/pnas.0905330106}
  {\bibfield  {journal} {\bibinfo  {journal} {Proceedings of the National
  Academy of Sciences}\ }\textbf {\bibinfo {volume} {106}},\ \bibinfo {pages}
  {15186} (\bibinfo {year} {2009})}\BibitemShut {NoStop}%
\bibitem [{\citenamefont {Cuny}\ \emph {et~al.}(2022)\citenamefont {Cuny},
  \citenamefont {Bertin},\ and\ \citenamefont
  {Mari}}]{cunyDynamicsMicrostructureAnisotropy2022}%
  \BibitemOpen
  \bibfield  {author} {\bibinfo {author} {\bibfnamefont {N.}~\bibnamefont
  {Cuny}}, \bibinfo {author} {\bibfnamefont {E.}~\bibnamefont {Bertin}},\ and\
  \bibinfo {author} {\bibfnamefont {R.}~\bibnamefont {Mari}},\ }\bibfield
  {title} {\bibinfo {title} {Dynamics of microstructure anisotropy and rheology
  of soft jammed suspensions},\ }\href {https://doi.org/10.1039/D1SM01345A}
  {\bibfield  {journal} {\bibinfo  {journal} {Soft Matter}\ }\textbf {\bibinfo
  {volume} {18}},\ \bibinfo {pages} {328} (\bibinfo {year} {2022})}\BibitemShut
  {NoStop}%
\bibitem [{\citenamefont {Seth}\ \emph {et~al.}(2011)\citenamefont {Seth},
  \citenamefont {Mohan}, \citenamefont {{Locatelli-Champagne}}, \citenamefont
  {Cloitre},\ and\ \citenamefont
  {Bonnecaze}}]{sethMicromechanicalModelPredict2011}%
  \BibitemOpen
  \bibfield  {author} {\bibinfo {author} {\bibfnamefont {J.~R.}\ \bibnamefont
  {Seth}}, \bibinfo {author} {\bibfnamefont {L.}~\bibnamefont {Mohan}},
  \bibinfo {author} {\bibfnamefont {C.}~\bibnamefont {{Locatelli-Champagne}}},
  \bibinfo {author} {\bibfnamefont {M.}~\bibnamefont {Cloitre}},\ and\ \bibinfo
  {author} {\bibfnamefont {R.~T.}\ \bibnamefont {Bonnecaze}},\ }\bibfield
  {title} {\bibinfo {title} {A micromechanical model to predict the flow of
  soft particle glasses},\ }\href {https://doi.org/10.1038/nmat3119} {\bibfield
   {journal} {\bibinfo  {journal} {Nature Materials}\ }\textbf {\bibinfo
  {volume} {10}},\ \bibinfo {pages} {838} (\bibinfo {year} {2011})},\ \Eprint
  {https://arxiv.org/abs/00641346} {00641346} \BibitemShut {NoStop}%
\bibitem [{\citenamefont {Cohen-Addad}\ \emph {et~al.}(2013)\citenamefont
  {Cohen-Addad}, \citenamefont {H{\"o}hler},\ and\ \citenamefont
  {Pitois}}]{cohen2013flow}%
  \BibitemOpen
  \bibfield  {author} {\bibinfo {author} {\bibfnamefont {S.}~\bibnamefont
  {Cohen-Addad}}, \bibinfo {author} {\bibfnamefont {R.}~\bibnamefont
  {H{\"o}hler}},\ and\ \bibinfo {author} {\bibfnamefont {O.}~\bibnamefont
  {Pitois}},\ }\bibfield  {title} {\bibinfo {title} {Flow in foams and flowing
  foams},\ }\href@noop {} {\bibfield  {journal} {\bibinfo  {journal} {Annual
  Review of Fluid Mechanics}\ }\textbf {\bibinfo {volume} {45}},\ \bibinfo
  {pages} {241} (\bibinfo {year} {2013})}\BibitemShut {NoStop}%
\bibitem [{\citenamefont {Sollich}\ \emph {et~al.}(1997)\citenamefont
  {Sollich}, \citenamefont {Lequeux}, \citenamefont {H{\'e}braud},\ and\
  \citenamefont {Cates}}]{sollich1997rheology}%
  \BibitemOpen
  \bibfield  {author} {\bibinfo {author} {\bibfnamefont {P.}~\bibnamefont
  {Sollich}}, \bibinfo {author} {\bibfnamefont {F.}~\bibnamefont {Lequeux}},
  \bibinfo {author} {\bibfnamefont {P.}~\bibnamefont {H{\'e}braud}},\ and\
  \bibinfo {author} {\bibfnamefont {M.~E.}\ \bibnamefont {Cates}},\ }\bibfield
  {title} {\bibinfo {title} {Rheology of soft glassy materials},\ }\href@noop
  {} {\bibfield  {journal} {\bibinfo  {journal} {Physical Review Letters}\
  }\textbf {\bibinfo {volume} {78}},\ \bibinfo {pages} {2020} (\bibinfo {year}
  {1997})}\BibitemShut {NoStop}%
\bibitem [{\citenamefont {Hebraud}\ and\ \citenamefont
  {Lequeux}(1998)}]{hebraud1998mode}%
  \BibitemOpen
  \bibfield  {author} {\bibinfo {author} {\bibfnamefont {P.}~\bibnamefont
  {Hebraud}}\ and\ \bibinfo {author} {\bibfnamefont {F.}~\bibnamefont
  {Lequeux}},\ }\bibfield  {title} {\bibinfo {title} {Mode-coupling theory for
  the pasty rheology of soft glassy materials},\ }\href@noop {} {\bibfield
  {journal} {\bibinfo  {journal} {Physical Review Letters}\ }\textbf {\bibinfo
  {volume} {81}},\ \bibinfo {pages} {2934} (\bibinfo {year}
  {1998})}\BibitemShut {NoStop}%
\bibitem [{\citenamefont {Langer}(2008)}]{langer2008shear}%
  \BibitemOpen
  \bibfield  {author} {\bibinfo {author} {\bibfnamefont {J.~S.}\ \bibnamefont
  {Langer}},\ }\bibfield  {title} {\bibinfo {title} {Shear-transformation-zone
  theory of plastic deformation near the glass transition},\ }\href@noop {}
  {\bibfield  {journal} {\bibinfo  {journal} {Physical Review E}\ }\textbf
  {\bibinfo {volume} {77}},\ \bibinfo {pages} {021502} (\bibinfo {year}
  {2008})}\BibitemShut {NoStop}%
\bibitem [{\citenamefont {Bocquet}\ \emph {et~al.}(2009)\citenamefont
  {Bocquet}, \citenamefont {Colin},\ and\ \citenamefont
  {Ajdari}}]{bocquet2009kinetic}%
  \BibitemOpen
  \bibfield  {author} {\bibinfo {author} {\bibfnamefont {L.}~\bibnamefont
  {Bocquet}}, \bibinfo {author} {\bibfnamefont {A.}~\bibnamefont {Colin}},\
  and\ \bibinfo {author} {\bibfnamefont {A.}~\bibnamefont {Ajdari}},\
  }\bibfield  {title} {\bibinfo {title} {Kinetic theory of plastic flow in soft
  glassy materials},\ }\href@noop {} {\bibfield  {journal} {\bibinfo  {journal}
  {Physical Review Letters}\ }\textbf {\bibinfo {volume} {103}},\ \bibinfo
  {pages} {036001} (\bibinfo {year} {2009})}\BibitemShut {NoStop}%
\bibitem [{\citenamefont {Moorcroft}\ and\ \citenamefont
  {Fielding}(2013)}]{moorcroft2013criteria}%
  \BibitemOpen
  \bibfield  {author} {\bibinfo {author} {\bibfnamefont {R.~L.}\ \bibnamefont
  {Moorcroft}}\ and\ \bibinfo {author} {\bibfnamefont {S.~M.}\ \bibnamefont
  {Fielding}},\ }\bibfield  {title} {\bibinfo {title} {Criteria for shear
  banding in time-dependent flows of complex fluids},\ }\href@noop {}
  {\bibfield  {journal} {\bibinfo  {journal} {Physical Review Letters}\
  }\textbf {\bibinfo {volume} {110}},\ \bibinfo {pages} {086001} (\bibinfo
  {year} {2013})}\BibitemShut {NoStop}%
\bibitem [{\citenamefont {Benzi}\ \emph {et~al.}(2016)\citenamefont {Benzi},
  \citenamefont {Sbragaglia}, \citenamefont {Bernaschi}, \citenamefont
  {Succi},\ and\ \citenamefont {Toschi}}]{benzi2016cooperativity}%
  \BibitemOpen
  \bibfield  {author} {\bibinfo {author} {\bibfnamefont {R.}~\bibnamefont
  {Benzi}}, \bibinfo {author} {\bibfnamefont {M.}~\bibnamefont {Sbragaglia}},
  \bibinfo {author} {\bibfnamefont {M.}~\bibnamefont {Bernaschi}}, \bibinfo
  {author} {\bibfnamefont {S.}~\bibnamefont {Succi}},\ and\ \bibinfo {author}
  {\bibfnamefont {F.}~\bibnamefont {Toschi}},\ }\bibfield  {title} {\bibinfo
  {title} {Cooperativity flows and shear-bandings: a statistical field theory
  approach},\ }\href@noop {} {\bibfield  {journal} {\bibinfo  {journal} {Soft
  Matter}\ }\textbf {\bibinfo {volume} {12}},\ \bibinfo {pages} {514} (\bibinfo
  {year} {2016})}\BibitemShut {NoStop}%
\bibitem [{\citenamefont {Nicolas}\ \emph {et~al.}(2014)\citenamefont
  {Nicolas}, \citenamefont {Martens}, \citenamefont {Bocquet},\ and\
  \citenamefont {Barrat}}]{nicolas2014universal}%
  \BibitemOpen
  \bibfield  {author} {\bibinfo {author} {\bibfnamefont {A.}~\bibnamefont
  {Nicolas}}, \bibinfo {author} {\bibfnamefont {K.}~\bibnamefont {Martens}},
  \bibinfo {author} {\bibfnamefont {L.}~\bibnamefont {Bocquet}},\ and\ \bibinfo
  {author} {\bibfnamefont {J.-L.}\ \bibnamefont {Barrat}},\ }\bibfield  {title}
  {\bibinfo {title} {Universal and non-universal features in coarse-grained
  models of flow in disordered solids},\ }\href@noop {} {\bibfield  {journal}
  {\bibinfo  {journal} {Soft matter}\ }\textbf {\bibinfo {volume} {10}},\
  \bibinfo {pages} {4648} (\bibinfo {year} {2014})}\BibitemShut {NoStop}%
\bibitem [{\citenamefont {Goyon}\ \emph {et~al.}(2008)\citenamefont {Goyon},
  \citenamefont {Colin}, \citenamefont {Ovarlez}, \citenamefont {Ajdari},\ and\
  \citenamefont {Bocquet}}]{goyon2008spatial}%
  \BibitemOpen
  \bibfield  {author} {\bibinfo {author} {\bibfnamefont {J.}~\bibnamefont
  {Goyon}}, \bibinfo {author} {\bibfnamefont {A.}~\bibnamefont {Colin}},
  \bibinfo {author} {\bibfnamefont {G.}~\bibnamefont {Ovarlez}}, \bibinfo
  {author} {\bibfnamefont {A.}~\bibnamefont {Ajdari}},\ and\ \bibinfo {author}
  {\bibfnamefont {L.}~\bibnamefont {Bocquet}},\ }\bibfield  {title} {\bibinfo
  {title} {Spatial cooperativity in soft glassy flows},\ }\href@noop {}
  {\bibfield  {journal} {\bibinfo  {journal} {Nature}\ }\textbf {\bibinfo
  {volume} {454}},\ \bibinfo {pages} {84} (\bibinfo {year} {2008})}\BibitemShut
  {NoStop}%
\bibitem [{\citenamefont {Benzi}\ \emph {et~al.}(2019)\citenamefont {Benzi},
  \citenamefont {Divoux}, \citenamefont {Barentin}, \citenamefont {Manneville},
  \citenamefont {Sbragaglia},\ and\ \citenamefont {Toschi}}]{benzi2019unified}%
  \BibitemOpen
  \bibfield  {author} {\bibinfo {author} {\bibfnamefont {R.}~\bibnamefont
  {Benzi}}, \bibinfo {author} {\bibfnamefont {T.}~\bibnamefont {Divoux}},
  \bibinfo {author} {\bibfnamefont {C.}~\bibnamefont {Barentin}}, \bibinfo
  {author} {\bibfnamefont {S.}~\bibnamefont {Manneville}}, \bibinfo {author}
  {\bibfnamefont {M.}~\bibnamefont {Sbragaglia}},\ and\ \bibinfo {author}
  {\bibfnamefont {F.}~\bibnamefont {Toschi}},\ }\bibfield  {title} {\bibinfo
  {title} {Unified theoretical and experimental view on transient shear
  banding},\ }\href@noop {} {\bibfield  {journal} {\bibinfo  {journal}
  {Physical Review Letters}\ }\textbf {\bibinfo {volume} {123}},\ \bibinfo
  {pages} {248001} (\bibinfo {year} {2019})}\BibitemShut {NoStop}%
\bibitem [{\citenamefont {Budrikis}\ \emph {et~al.}(2017)\citenamefont
  {Budrikis}, \citenamefont {Castellanos}, \citenamefont {Sandfeld},
  \citenamefont {Zaiser},\ and\ \citenamefont
  {Zapperi}}]{budrikisUniversalFeaturesAmorphous2017}%
  \BibitemOpen
  \bibfield  {author} {\bibinfo {author} {\bibfnamefont {Z.}~\bibnamefont
  {Budrikis}}, \bibinfo {author} {\bibfnamefont {D.~F.}\ \bibnamefont
  {Castellanos}}, \bibinfo {author} {\bibfnamefont {S.}~\bibnamefont
  {Sandfeld}}, \bibinfo {author} {\bibfnamefont {M.}~\bibnamefont {Zaiser}},\
  and\ \bibinfo {author} {\bibfnamefont {S.}~\bibnamefont {Zapperi}},\
  }\bibfield  {title} {\bibinfo {title} {Universal features of amorphous
  plasticity},\ }\href {https://doi.org/10.1038/ncomms15928} {\bibfield
  {journal} {\bibinfo  {journal} {Nature Communications}\ }\textbf {\bibinfo
  {volume} {8}},\ \bibinfo {pages} {15928} (\bibinfo {year} {2017})},\ \Eprint
  {https://arxiv.org/abs/1511.06229} {1511.06229} \BibitemShut {NoStop}%
\bibitem [{\citenamefont {Liu}\ \emph {et~al.}(2018)\citenamefont {Liu},
  \citenamefont {Ferrero}, \citenamefont {Martens},\ and\ \citenamefont
  {Barrat}}]{liu2018creep}%
  \BibitemOpen
  \bibfield  {author} {\bibinfo {author} {\bibfnamefont {C.}~\bibnamefont
  {Liu}}, \bibinfo {author} {\bibfnamefont {E.~E.}\ \bibnamefont {Ferrero}},
  \bibinfo {author} {\bibfnamefont {K.}~\bibnamefont {Martens}},\ and\ \bibinfo
  {author} {\bibfnamefont {J.-L.}\ \bibnamefont {Barrat}},\ }\bibfield  {title}
  {\bibinfo {title} {Creep dynamics of athermal amorphous materials: a
  mesoscopic approach},\ }\href@noop {} {\bibfield  {journal} {\bibinfo
  {journal} {Soft matter}\ }\textbf {\bibinfo {volume} {14}},\ \bibinfo {pages}
  {8306} (\bibinfo {year} {2018})}\BibitemShut {NoStop}%
\bibitem [{\citenamefont {Ozawa}\ \emph {et~al.}(2018)\citenamefont {Ozawa},
  \citenamefont {Berthier}, \citenamefont {Biroli}, \citenamefont {Rosso},\
  and\ \citenamefont {Tarjus}}]{ozawa2018random}%
  \BibitemOpen
  \bibfield  {author} {\bibinfo {author} {\bibfnamefont {M.}~\bibnamefont
  {Ozawa}}, \bibinfo {author} {\bibfnamefont {L.}~\bibnamefont {Berthier}},
  \bibinfo {author} {\bibfnamefont {G.}~\bibnamefont {Biroli}}, \bibinfo
  {author} {\bibfnamefont {A.}~\bibnamefont {Rosso}},\ and\ \bibinfo {author}
  {\bibfnamefont {G.}~\bibnamefont {Tarjus}},\ }\bibfield  {title} {\bibinfo
  {title} {Random critical point separates brittle and ductile yielding
  transitions in amorphous materials},\ }\href@noop {} {\bibfield  {journal}
  {\bibinfo  {journal} {Proceedings of the National Academy of Sciences}\
  }\textbf {\bibinfo {volume} {115}},\ \bibinfo {pages} {6656} (\bibinfo {year}
  {2018})}\BibitemShut {NoStop}%
\bibitem [{\citenamefont {Popovi{\'c}}\ \emph {et~al.}(2018)\citenamefont
  {Popovi{\'c}}, \citenamefont {de~Geus},\ and\ \citenamefont
  {Wyart}}]{popovic2018elastoplastic}%
  \BibitemOpen
  \bibfield  {author} {\bibinfo {author} {\bibfnamefont {M.}~\bibnamefont
  {Popovi{\'c}}}, \bibinfo {author} {\bibfnamefont {T.~W.}\ \bibnamefont
  {de~Geus}},\ and\ \bibinfo {author} {\bibfnamefont {M.}~\bibnamefont
  {Wyart}},\ }\bibfield  {title} {\bibinfo {title} {Elastoplastic description
  of sudden failure in athermal amorphous materials during quasistatic
  loading},\ }\href@noop {} {\bibfield  {journal} {\bibinfo  {journal}
  {Physical Review E}\ }\textbf {\bibinfo {volume} {98}},\ \bibinfo {pages}
  {040901} (\bibinfo {year} {2018})}\BibitemShut {NoStop}%
\bibitem [{\citenamefont {Patinet}\ \emph {et~al.}(2020)\citenamefont
  {Patinet}, \citenamefont {Barbot}, \citenamefont {Lerbinger}, \citenamefont
  {Vandembroucq},\ and\ \citenamefont {Lema{\^\i}tre}}]{patinet2020origin}%
  \BibitemOpen
  \bibfield  {author} {\bibinfo {author} {\bibfnamefont {S.}~\bibnamefont
  {Patinet}}, \bibinfo {author} {\bibfnamefont {A.}~\bibnamefont {Barbot}},
  \bibinfo {author} {\bibfnamefont {M.}~\bibnamefont {Lerbinger}}, \bibinfo
  {author} {\bibfnamefont {D.}~\bibnamefont {Vandembroucq}},\ and\ \bibinfo
  {author} {\bibfnamefont {A.}~\bibnamefont {Lema{\^\i}tre}},\ }\bibfield
  {title} {\bibinfo {title} {Origin of the bauschinger effect in amorphous
  solids},\ }\href@noop {} {\bibfield  {journal} {\bibinfo  {journal} {Physical
  Review Letters}\ }\textbf {\bibinfo {volume} {124}},\ \bibinfo {pages}
  {205503} (\bibinfo {year} {2020})}\BibitemShut {NoStop}%
\bibitem [{\citenamefont {Ruan}\ \emph {et~al.}(2022)\citenamefont {Ruan},
  \citenamefont {Patinet},\ and\ \citenamefont {Falk}}]{ruan2022predicting}%
  \BibitemOpen
  \bibfield  {author} {\bibinfo {author} {\bibfnamefont {D.}~\bibnamefont
  {Ruan}}, \bibinfo {author} {\bibfnamefont {S.}~\bibnamefont {Patinet}},\ and\
  \bibinfo {author} {\bibfnamefont {M.~L.}\ \bibnamefont {Falk}},\ }\bibfield
  {title} {\bibinfo {title} {Predicting plastic events and quantifying the
  local yield surface in 3d model glasses},\ }\href@noop {} {\bibfield
  {journal} {\bibinfo  {journal} {Journal of the Mechanics and Physics of
  Solids}\ }\textbf {\bibinfo {volume} {158}},\ \bibinfo {pages} {104671}
  (\bibinfo {year} {2022})}\BibitemShut {NoStop}%
\bibitem [{\citenamefont {Puosi}\ \emph {et~al.}(2015)\citenamefont {Puosi},
  \citenamefont {Olivier},\ and\ \citenamefont {Martens}}]{puosi2015probing}%
  \BibitemOpen
  \bibfield  {author} {\bibinfo {author} {\bibfnamefont {F.}~\bibnamefont
  {Puosi}}, \bibinfo {author} {\bibfnamefont {J.}~\bibnamefont {Olivier}},\
  and\ \bibinfo {author} {\bibfnamefont {K.}~\bibnamefont {Martens}},\
  }\bibfield  {title} {\bibinfo {title} {Probing relevant ingredients in
  mean-field approaches for the athermal rheology of yield stress materials},\
  }\href {https://doi.org/10.1039/C5SM01694K} {\bibfield  {journal} {\bibinfo
  {journal} {Soft Matter}\ }\textbf {\bibinfo {volume} {11}},\ \bibinfo {pages}
  {7639} (\bibinfo {year} {2015})},\ \Eprint {https://arxiv.org/abs/1501.04574}
  {1501.04574} \BibitemShut {NoStop}%
\bibitem [{\citenamefont {Patinet}\ \emph {et~al.}(2016)\citenamefont
  {Patinet}, \citenamefont {Vandembroucq},\ and\ \citenamefont
  {Falk}}]{patinet2016connecting}%
  \BibitemOpen
  \bibfield  {author} {\bibinfo {author} {\bibfnamefont {S.}~\bibnamefont
  {Patinet}}, \bibinfo {author} {\bibfnamefont {D.}~\bibnamefont
  {Vandembroucq}},\ and\ \bibinfo {author} {\bibfnamefont {M.~L.}\ \bibnamefont
  {Falk}},\ }\bibfield  {title} {\bibinfo {title} {Connecting local yield
  stresses with plastic activity in amorphous solids},\ }\href@noop {}
  {\bibfield  {journal} {\bibinfo  {journal} {Physical Review Letters}\
  }\textbf {\bibinfo {volume} {117}},\ \bibinfo {pages} {045501} (\bibinfo
  {year} {2016})}\BibitemShut {NoStop}%
\bibitem [{\citenamefont {Liu}\ \emph {et~al.}(2021)\citenamefont {Liu},
  \citenamefont {Dutta}, \citenamefont {Chaudhuri},\ and\ \citenamefont
  {Martens}}]{liuElastoplasticApproachBased2021}%
  \BibitemOpen
  \bibfield  {author} {\bibinfo {author} {\bibfnamefont {C.}~\bibnamefont
  {Liu}}, \bibinfo {author} {\bibfnamefont {S.}~\bibnamefont {Dutta}}, \bibinfo
  {author} {\bibfnamefont {P.}~\bibnamefont {Chaudhuri}},\ and\ \bibinfo
  {author} {\bibfnamefont {K.}~\bibnamefont {Martens}},\ }\bibfield  {title}
  {\bibinfo {title} {Elastoplastic {{Approach Based}} on {{Microscopic
  Insights}} for the {{Steady State}} and {{Transient Dynamics}} of {{Sheared
  Disordered Solids}}},\ }\href
  {https://doi.org/10.1103/PhysRevLett.126.138005} {\bibfield  {journal}
  {\bibinfo  {journal} {Physical Review Letters}\ }\textbf {\bibinfo {volume}
  {126}},\ \bibinfo {pages} {138005} (\bibinfo {year} {2021})},\ \Eprint
  {https://arxiv.org/abs/2007.07162} {arXiv:2007.07162 [cond-mat]} \BibitemShut
  {NoStop}%
\bibitem [{\citenamefont {Castellanos}\ \emph {et~al.}(2021)\citenamefont
  {Castellanos}, \citenamefont {Roux},\ and\ \citenamefont
  {Patinet}}]{castellanos2021insights}%
  \BibitemOpen
  \bibfield  {author} {\bibinfo {author} {\bibfnamefont {D.~F.}\ \bibnamefont
  {Castellanos}}, \bibinfo {author} {\bibfnamefont {S.}~\bibnamefont {Roux}},\
  and\ \bibinfo {author} {\bibfnamefont {S.}~\bibnamefont {Patinet}},\
  }\bibfield  {title} {\bibinfo {title} {Insights from the quantitative
  calibration of an elasto-plastic model from a lennard-jones atomic glass},\
  }\href@noop {} {\bibfield  {journal} {\bibinfo  {journal} {Comptes Rendus.
  Physique}\ }\textbf {\bibinfo {volume} {22}},\ \bibinfo {pages} {1} (\bibinfo
  {year} {2021})}\BibitemShut {NoStop}%
\bibitem [{\citenamefont {Vincent}\ and\ \citenamefont
  {Schurtenberger}(2011)}]{vincent2011work}%
  \BibitemOpen
  \bibfield  {author} {\bibinfo {author} {\bibfnamefont {R.~R.}\ \bibnamefont
  {Vincent}}\ and\ \bibinfo {author} {\bibfnamefont {P.}~\bibnamefont
  {Schurtenberger}},\ }\bibfield  {title} {\bibinfo {title} {Work hardening of
  soft glassy materials, or a metallurgist’s view of peanut butter},\
  }\href@noop {} {\bibfield  {journal} {\bibinfo  {journal} {Soft Matter}\
  }\textbf {\bibinfo {volume} {7}},\ \bibinfo {pages} {1635} (\bibinfo {year}
  {2011})}\BibitemShut {NoStop}%
\bibitem [{\citenamefont {Dimitriou}\ \emph {et~al.}(2013)\citenamefont
  {Dimitriou}, \citenamefont {Ewoldt},\ and\ \citenamefont
  {McKinley}}]{dimitriou2013describing}%
  \BibitemOpen
  \bibfield  {author} {\bibinfo {author} {\bibfnamefont {C.~J.}\ \bibnamefont
  {Dimitriou}}, \bibinfo {author} {\bibfnamefont {R.~H.}\ \bibnamefont
  {Ewoldt}},\ and\ \bibinfo {author} {\bibfnamefont {G.~H.}\ \bibnamefont
  {McKinley}},\ }\bibfield  {title} {\bibinfo {title} {Describing and
  prescribing the constitutive response of yield stress fluids using large
  amplitude oscillatory shear stress (laostress)},\ }\href@noop {} {\bibfield
  {journal} {\bibinfo  {journal} {Journal of Rheology}\ }\textbf {\bibinfo
  {volume} {57}},\ \bibinfo {pages} {27} (\bibinfo {year} {2013})}\BibitemShut
  {NoStop}%
\bibitem [{\citenamefont {Bhattacharjee}(2015)}]{bhattacharjee2015stress}%
  \BibitemOpen
  \bibfield  {author} {\bibinfo {author} {\bibfnamefont {A.~K.}\ \bibnamefont
  {Bhattacharjee}},\ }\bibfield  {title} {\bibinfo {title} {Stress--structure
  relation in dense colloidal melts under forward and instantaneous reversal of
  the shear},\ }\href@noop {} {\bibfield  {journal} {\bibinfo  {journal} {Soft
  Matter}\ }\textbf {\bibinfo {volume} {11}},\ \bibinfo {pages} {5697}
  (\bibinfo {year} {2015})}\BibitemShut {NoStop}%
\bibitem [{\citenamefont {Dimitriou}\ and\ \citenamefont
  {McKinley}(2019)}]{dimitriou2019canonical}%
  \BibitemOpen
  \bibfield  {author} {\bibinfo {author} {\bibfnamefont {C.~J.}\ \bibnamefont
  {Dimitriou}}\ and\ \bibinfo {author} {\bibfnamefont {G.~H.}\ \bibnamefont
  {McKinley}},\ }\bibfield  {title} {\bibinfo {title} {A canonical framework
  for modeling elasto-viscoplasticity in complex fluids},\ }\href@noop {}
  {\bibfield  {journal} {\bibinfo  {journal} {J. Non-Newt. Fluid Mech.}\
  }\textbf {\bibinfo {volume} {265}},\ \bibinfo {pages} {116} (\bibinfo {year}
  {2019})}\BibitemShut {NoStop}%
\bibitem [{\citenamefont {Deboeuf}\ \emph {et~al.}(2022)\citenamefont
  {Deboeuf}, \citenamefont {Duclou{\'e}}, \citenamefont {Lenoir},\ and\
  \citenamefont {Ovarlez}}]{deboeufMechanismStrainHardening2022}%
  \BibitemOpen
  \bibfield  {author} {\bibinfo {author} {\bibfnamefont {S.}~\bibnamefont
  {Deboeuf}}, \bibinfo {author} {\bibfnamefont {L.}~\bibnamefont
  {Duclou{\'e}}}, \bibinfo {author} {\bibfnamefont {N.}~\bibnamefont
  {Lenoir}},\ and\ \bibinfo {author} {\bibfnamefont {G.}~\bibnamefont
  {Ovarlez}},\ }\bibfield  {title} {\bibinfo {title} {A mechanism of strain
  hardening and {{Bauschinger}} effect: Shear-history-dependent microstructure
  of elasto-plastic suspensions},\ }\href {https://doi.org/10.1039/D2SM00910B}
  {\bibfield  {journal} {\bibinfo  {journal} {Soft Matter}\ }\textbf {\bibinfo
  {volume} {18}},\ \bibinfo {pages} {8756} (\bibinfo {year} {2022})},\ \Eprint
  {https://arxiv.org/abs/03703343} {03703343} \BibitemShut {NoStop}%
\bibitem [{\citenamefont {Beyer}\ \emph {et~al.}(2025)\citenamefont {Beyer},
  \citenamefont {Engel},\ and\ \citenamefont
  {H{\'e}braud}}]{beyerShearinducedMechanicalAnisotropy2025}%
  \BibitemOpen
  \bibfield  {author} {\bibinfo {author} {\bibfnamefont {N.}~\bibnamefont
  {Beyer}}, \bibinfo {author} {\bibfnamefont {L.}~\bibnamefont {Engel}},\ and\
  \bibinfo {author} {\bibfnamefont {P.}~\bibnamefont {H{\'e}braud}},\
  }\bibfield  {title} {\bibinfo {title} {Shear-induced mechanical anisotropy in
  concentrated emulsions : A {{Pump Probe Mechanical Spectroscopy}} study}}
  (\bibinfo {year} {2025}),\ \bibinfo {note} {hal-05142986}\BibitemShut
  {NoStop}%
\bibitem [{\citenamefont {Seeger}\ \emph {et~al.}(1957)\citenamefont {Seeger},
  \citenamefont {Diehl}, \citenamefont {Mader},\ and\ \citenamefont
  {Rebstock}}]{seegerWorkhardeningWorksofteningFacecentred1957}%
  \BibitemOpen
  \bibfield  {author} {\bibinfo {author} {\bibfnamefont {A.}~\bibnamefont
  {Seeger}}, \bibinfo {author} {\bibfnamefont {J.}~\bibnamefont {Diehl}},
  \bibinfo {author} {\bibfnamefont {S.}~\bibnamefont {Mader}},\ and\ \bibinfo
  {author} {\bibfnamefont {H.}~\bibnamefont {Rebstock}},\ }\bibfield  {title}
  {\bibinfo {title} {Work-hardening and work-softening of face-centred cubic
  metal crystals},\ }\href {https://doi.org/10.1080/14786435708243823}
  {\bibfield  {journal} {\bibinfo  {journal} {The Philosophical Magazine: A
  Journal of Theoretical Experimental and Applied Physics}\ }\textbf {\bibinfo
  {volume} {2}},\ \bibinfo {pages} {323} (\bibinfo {year} {1957})}\BibitemShut
  {NoStop}%
\bibitem [{\citenamefont {Hirsch}(1964)}]{hirsch1964theory}%
  \BibitemOpen
  \bibfield  {author} {\bibinfo {author} {\bibfnamefont {P.}~\bibnamefont
  {Hirsch}},\ }\bibfield  {title} {\bibinfo {title} {A theory of linear strain
  hardening in crystals},\ }\href@noop {} {\bibfield  {journal} {\bibinfo
  {journal} {Discussions of the Faraday Society}\ }\textbf {\bibinfo {volume}
  {38}},\ \bibinfo {pages} {111} (\bibinfo {year} {1964})}\BibitemShut
  {NoStop}%
\bibitem [{\citenamefont {Hoy}\ and\ \citenamefont
  {Robbins}(2008)}]{hoy2008strain}%
  \BibitemOpen
  \bibfield  {author} {\bibinfo {author} {\bibfnamefont {R.~S.}\ \bibnamefont
  {Hoy}}\ and\ \bibinfo {author} {\bibfnamefont {M.~O.}\ \bibnamefont
  {Robbins}},\ }\bibfield  {title} {\bibinfo {title} {Strain hardening of
  polymer glasses: Entanglements, energetics, and plasticity},\ }\href@noop {}
  {\bibfield  {journal} {\bibinfo  {journal} {Phys. Rev. E}\ }\textbf {\bibinfo
  {volume} {77}},\ \bibinfo {pages} {031801} (\bibinfo {year}
  {2008})}\BibitemShut {NoStop}%
\bibitem [{\citenamefont {Ames}\ \emph {et~al.}(2009)\citenamefont {Ames},
  \citenamefont {Srivastava}, \citenamefont {Chester},\ and\ \citenamefont
  {Anand}}]{amesThermomechanicallyCoupledTheory2009}%
  \BibitemOpen
  \bibfield  {author} {\bibinfo {author} {\bibfnamefont {N.~M.}\ \bibnamefont
  {Ames}}, \bibinfo {author} {\bibfnamefont {V.}~\bibnamefont {Srivastava}},
  \bibinfo {author} {\bibfnamefont {S.~A.}\ \bibnamefont {Chester}},\ and\
  \bibinfo {author} {\bibfnamefont {L.}~\bibnamefont {Anand}},\ }\bibfield
  {title} {\bibinfo {title} {A thermo-mechanically coupled theory for large
  deformations of amorphous polymers. {{Part II}}: {{Applications}}},\ }\href
  {https://doi.org/10.1016/j.ijplas.2008.11.005} {\bibfield  {journal}
  {\bibinfo  {journal} {International Journal of Plasticity}\ }\textbf
  {\bibinfo {volume} {25}},\ \bibinfo {pages} {1495} (\bibinfo {year}
  {2009})}\BibitemShut {NoStop}%
\bibitem [{\citenamefont {Anand}\ \emph {et~al.}(2009)\citenamefont {Anand},
  \citenamefont {Ames}, \citenamefont {Srivastava},\ and\ \citenamefont
  {Chester}}]{anandThermomechanicallyCoupledTheory2009}%
  \BibitemOpen
  \bibfield  {author} {\bibinfo {author} {\bibfnamefont {L.}~\bibnamefont
  {Anand}}, \bibinfo {author} {\bibfnamefont {N.~M.}\ \bibnamefont {Ames}},
  \bibinfo {author} {\bibfnamefont {V.}~\bibnamefont {Srivastava}},\ and\
  \bibinfo {author} {\bibfnamefont {S.~A.}\ \bibnamefont {Chester}},\
  }\bibfield  {title} {\bibinfo {title} {A thermo-mechanically coupled theory
  for large deformations of amorphous polymers. {{Part I}}: {{Formulation}}},\
  }\href {https://doi.org/10.1016/j.ijplas.2008.11.004} {\bibfield  {journal}
  {\bibinfo  {journal} {International Journal of Plasticity}\ }\textbf
  {\bibinfo {volume} {25}},\ \bibinfo {pages} {1474} (\bibinfo {year}
  {2009})}\BibitemShut {NoStop}%
\bibitem [{\citenamefont {Pan}\ \emph {et~al.}(2020)\citenamefont {Pan},
  \citenamefont {Ivanov}, \citenamefont {Zhou}, \citenamefont {Li},\ and\
  \citenamefont {Greer}}]{pan2020strain}%
  \BibitemOpen
  \bibfield  {author} {\bibinfo {author} {\bibfnamefont {J.}~\bibnamefont
  {Pan}}, \bibinfo {author} {\bibfnamefont {Y.~P.}\ \bibnamefont {Ivanov}},
  \bibinfo {author} {\bibfnamefont {W.}~\bibnamefont {Zhou}}, \bibinfo {author}
  {\bibfnamefont {Y.}~\bibnamefont {Li}},\ and\ \bibinfo {author}
  {\bibfnamefont {A.}~\bibnamefont {Greer}},\ }\bibfield  {title} {\bibinfo
  {title} {Strain-hardening and suppression of shear-banding in rejuvenated
  bulk metallic glass},\ }\href@noop {} {\bibfield  {journal} {\bibinfo
  {journal} {Nature}\ }\textbf {\bibinfo {volume} {578}},\ \bibinfo {pages}
  {559} (\bibinfo {year} {2020})},\ \Eprint {https://arxiv.org/abs/302974}
  {302974} \BibitemShut {NoStop}%
\bibitem [{\citenamefont {Li}\ \emph {et~al.}(2001)\citenamefont {Li},
  \citenamefont {Wang}, \citenamefont {Wu} \emph {et~al.}}]{li2001tensile}%
  \BibitemOpen
  \bibfield  {author} {\bibinfo {author} {\bibfnamefont {V.~C.}\ \bibnamefont
  {Li}}, \bibinfo {author} {\bibfnamefont {S.}~\bibnamefont {Wang}}, \bibinfo
  {author} {\bibfnamefont {C.}~\bibnamefont {Wu}}, \emph {et~al.},\ }\bibfield
  {title} {\bibinfo {title} {Tensile strain-hardening behavior of polyvinyl
  alcohol engineered cementitious composite (pva-ecc)},\ }\href@noop {}
  {\bibfield  {journal} {\bibinfo  {journal} {ACI Mater. Journal-American
  Concrete Institute}\ }\textbf {\bibinfo {volume} {98}},\ \bibinfo {pages}
  {483} (\bibinfo {year} {2001})}\BibitemShut {NoStop}%
\bibitem [{\citenamefont {Weigandt}\ \emph {et~al.}(2011)\citenamefont
  {Weigandt}, \citenamefont {Porcar},\ and\ \citenamefont
  {Pozzo}}]{weigandt2011situ}%
  \BibitemOpen
  \bibfield  {author} {\bibinfo {author} {\bibfnamefont {K.~M.}\ \bibnamefont
  {Weigandt}}, \bibinfo {author} {\bibfnamefont {L.}~\bibnamefont {Porcar}},\
  and\ \bibinfo {author} {\bibfnamefont {D.~C.}\ \bibnamefont {Pozzo}},\
  }\bibfield  {title} {\bibinfo {title} {In situ neutron scattering study of
  structural transitions in fibrin networks under shear deformation},\
  }\href@noop {} {\bibfield  {journal} {\bibinfo  {journal} {Soft Matter}\
  }\textbf {\bibinfo {volume} {7}},\ \bibinfo {pages} {9992} (\bibinfo {year}
  {2011})}\BibitemShut {NoStop}%
\bibitem [{\citenamefont {Groot}\ \emph {et~al.}(1996)\citenamefont {Groot},
  \citenamefont {Bot},\ and\ \citenamefont {Agterof}}]{groot1996molecular}%
  \BibitemOpen
  \bibfield  {author} {\bibinfo {author} {\bibfnamefont {R.~D.}\ \bibnamefont
  {Groot}}, \bibinfo {author} {\bibfnamefont {A.}~\bibnamefont {Bot}},\ and\
  \bibinfo {author} {\bibfnamefont {W.~G.}\ \bibnamefont {Agterof}},\
  }\bibfield  {title} {\bibinfo {title} {Molecular theory of strain hardening
  of a polymer gel: application to gelatin},\ }\href@noop {} {\bibfield
  {journal} {\bibinfo  {journal} {J. Chem. Phys.}\ }\textbf {\bibinfo {volume}
  {104}},\ \bibinfo {pages} {9202} (\bibinfo {year} {1996})}\BibitemShut
  {NoStop}%
\bibitem [{\citenamefont {An}\ \emph {et~al.}(2012)\citenamefont {An},
  \citenamefont {Picken},\ and\ \citenamefont {Mendes}}]{an2012direct}%
  \BibitemOpen
  \bibfield  {author} {\bibinfo {author} {\bibfnamefont {H.-N.}\ \bibnamefont
  {An}}, \bibinfo {author} {\bibfnamefont {S.~J.}\ \bibnamefont {Picken}},\
  and\ \bibinfo {author} {\bibfnamefont {E.}~\bibnamefont {Mendes}},\
  }\bibfield  {title} {\bibinfo {title} {Direct observation of particle
  rearrangement during cyclic stress hardening of magnetorheological gels},\
  }\href@noop {} {\bibfield  {journal} {\bibinfo  {journal} {Soft Matter}\
  }\textbf {\bibinfo {volume} {8}},\ \bibinfo {pages} {11995} (\bibinfo {year}
  {2012})}\BibitemShut {NoStop}%
\bibitem [{\citenamefont {Laurati}\ \emph {et~al.}(2014)\citenamefont
  {Laurati}, \citenamefont {Egelhaaf},\ and\ \citenamefont
  {Petekidis}}]{laurati2014plastic}%
  \BibitemOpen
  \bibfield  {author} {\bibinfo {author} {\bibfnamefont {M.}~\bibnamefont
  {Laurati}}, \bibinfo {author} {\bibfnamefont {S.}~\bibnamefont {Egelhaaf}},\
  and\ \bibinfo {author} {\bibfnamefont {G.}~\bibnamefont {Petekidis}},\
  }\bibfield  {title} {\bibinfo {title} {Plastic rearrangements in colloidal
  gels investigated by laos and ls-echo},\ }\href@noop {} {\bibfield  {journal}
  {\bibinfo  {journal} {J. Rheo.}\ }\textbf {\bibinfo {volume} {58}},\ \bibinfo
  {pages} {1395} (\bibinfo {year} {2014})}\BibitemShut {NoStop}%
\bibitem [{\citenamefont {Buckley}\ and\ \citenamefont
  {Entwistle}(1956)}]{Buckley56}%
  \BibitemOpen
  \bibfield  {author} {\bibinfo {author} {\bibfnamefont {S.}~\bibnamefont
  {Buckley}}\ and\ \bibinfo {author} {\bibfnamefont {K.}~\bibnamefont
  {Entwistle}},\ }\bibfield  {title} {\bibinfo {title} {The bauschinger effect
  in super-pure aluminum single crystals and polycrystals},\ }\href
  {https://doi.org/https://doi.org/10.1016/0001-6160(56)90023-2} {\bibfield
  {journal} {\bibinfo  {journal} {Acta Metall.}\ }\textbf {\bibinfo {volume}
  {4}},\ \bibinfo {pages} {352} (\bibinfo {year} {1956})}\BibitemShut {NoStop}%
\bibitem [{\citenamefont {Asaro}(1975)}]{Asaro75}%
  \BibitemOpen
  \bibfield  {author} {\bibinfo {author} {\bibfnamefont {R.}~\bibnamefont
  {Asaro}},\ }\bibfield  {title} {\bibinfo {title} {Elastic-plastic memory and
  kinematic-type hardening},\ }\href
  {https://doi.org/https://doi.org/10.1016/0001-6160(75)90044-9} {\bibfield
  {journal} {\bibinfo  {journal} {Acta Metall.}\ }\textbf {\bibinfo {volume}
  {23}},\ \bibinfo {pages} {1255} (\bibinfo {year} {1975})}\BibitemShut
  {NoStop}%
\bibitem [{\citenamefont {Frank}(1980)}]{Frank80}%
  \BibitemOpen
  \bibfield  {author} {\bibinfo {author} {\bibfnamefont {F.~C.}\ \bibnamefont
  {Frank}},\ }\bibfield  {title} {\bibinfo {title} {The frank—read source},\
  }\href@noop {} {\bibfield  {journal} {\bibinfo  {journal} {Proc. R. Soc.
  Lond. A}\ }\textbf {\bibinfo {volume} {371}},\ \bibinfo {pages} {136–138}
  (\bibinfo {year} {1980})}\BibitemShut {NoStop}%
\bibitem [{\citenamefont {Karmakar}\ \emph {et~al.}(2010)\citenamefont
  {Karmakar}, \citenamefont {Lerner},\ and\ \citenamefont
  {Procaccia}}]{Karmakar10}%
  \BibitemOpen
  \bibfield  {author} {\bibinfo {author} {\bibfnamefont {S.}~\bibnamefont
  {Karmakar}}, \bibinfo {author} {\bibfnamefont {E.}~\bibnamefont {Lerner}},\
  and\ \bibinfo {author} {\bibfnamefont {I.}~\bibnamefont {Procaccia}},\
  }\bibfield  {title} {\bibinfo {title} {Plasticity-induced anisotropy in
  amorphous solids: The bauschinger effect},\ }\href@noop {} {\bibfield
  {journal} {\bibinfo  {journal} {Phys. Rev. E}\ }\textbf {\bibinfo {volume}
  {82}},\ \bibinfo {pages} {026104} (\bibinfo {year} {2010})}\BibitemShut
  {NoStop}%
\bibitem [{\citenamefont {Edera}\ \emph {et~al.}(2025)\citenamefont {Edera},
  \citenamefont {Bantawa}, \citenamefont {Aime}, \citenamefont {Bonnecaze},\
  and\ \citenamefont {Cloitre}}]{ederaTuningResidualStress2025}%
  \BibitemOpen
  \bibfield  {author} {\bibinfo {author} {\bibfnamefont {P.}~\bibnamefont
  {Edera}}, \bibinfo {author} {\bibfnamefont {M.}~\bibnamefont {Bantawa}},
  \bibinfo {author} {\bibfnamefont {S.}~\bibnamefont {Aime}}, \bibinfo {author}
  {\bibfnamefont {R.~T.}\ \bibnamefont {Bonnecaze}},\ and\ \bibinfo {author}
  {\bibfnamefont {M.}~\bibnamefont {Cloitre}},\ }\bibfield  {title} {\bibinfo
  {title} {Mechanical tuning of residual stress, memory, and aging in soft
  glassy materials},\ }\href {https://doi.org/10.1103/PhysRevX.15.011043}
  {\bibfield  {journal} {\bibinfo  {journal} {Phys. Rev. X}\ }\textbf {\bibinfo
  {volume} {15}},\ \bibinfo {pages} {011043} (\bibinfo {year}
  {2025})}\BibitemShut {NoStop}%
\bibitem [{\citenamefont {Blanc}\ \emph {et~al.}(2023)\citenamefont {Blanc},
  \citenamefont {Peters}, \citenamefont {Gillissen}, \citenamefont {Cates},
  \citenamefont {Bosio}, \citenamefont {Benarroche},\ and\ \citenamefont
  {Mari}}]{blancRheologyDenseSuspensions2023}%
  \BibitemOpen
  \bibfield  {author} {\bibinfo {author} {\bibfnamefont {F.}~\bibnamefont
  {Blanc}}, \bibinfo {author} {\bibfnamefont {F.}~\bibnamefont {Peters}},
  \bibinfo {author} {\bibfnamefont {J.~J.~J.}\ \bibnamefont {Gillissen}},
  \bibinfo {author} {\bibfnamefont {M.~E.}\ \bibnamefont {Cates}}, \bibinfo
  {author} {\bibfnamefont {S.}~\bibnamefont {Bosio}}, \bibinfo {author}
  {\bibfnamefont {C.}~\bibnamefont {Benarroche}},\ and\ \bibinfo {author}
  {\bibfnamefont {R.}~\bibnamefont {Mari}},\ }\bibfield  {title} {\bibinfo
  {title} {Rheology of {{Dense Suspensions}} under {{Shear Rotation}}},\ }\href
  {https://doi.org/10.1103/PhysRevLett.130.118202} {\bibfield  {journal}
  {\bibinfo  {journal} {Physical Review Letters}\ }\textbf {\bibinfo {volume}
  {130}},\ \bibinfo {pages} {118202} (\bibinfo {year} {2023})},\ \Eprint
  {https://arxiv.org/abs/03837785} {03837785} \BibitemShut {NoStop}%
\bibitem [{\citenamefont {Piau}(2007)}]{piau2007carbopol}%
  \BibitemOpen
  \bibfield  {author} {\bibinfo {author} {\bibfnamefont {J.-M.}\ \bibnamefont
  {Piau}},\ }\bibfield  {title} {\bibinfo {title} {Carbopol gels:
  Elastoviscoplastic and slippery glasses made of individual swollen sponges:
  Meso-and macroscopic properties, constitutive equations and scaling laws},\
  }\href {https://doi.org/10.1016/j.jnnfm.2007.02.011} {\bibfield  {journal}
  {\bibinfo  {journal} {Journal of Non-Newtonian Fluid Mechanics}\ }\textbf
  {\bibinfo {volume} {144}},\ \bibinfo {pages} {1} (\bibinfo {year}
  {2007})}\BibitemShut {NoStop}%
\bibitem [{\citenamefont {Mahaut}\ \emph {et~al.}(2008)\citenamefont {Mahaut},
  \citenamefont {Chateau}, \citenamefont {Coussot},\ and\ \citenamefont
  {Ovarlez}}]{mahaut2008yield}%
  \BibitemOpen
  \bibfield  {author} {\bibinfo {author} {\bibfnamefont {F.}~\bibnamefont
  {Mahaut}}, \bibinfo {author} {\bibfnamefont {X.}~\bibnamefont {Chateau}},
  \bibinfo {author} {\bibfnamefont {P.}~\bibnamefont {Coussot}},\ and\ \bibinfo
  {author} {\bibfnamefont {G.}~\bibnamefont {Ovarlez}},\ }\bibfield  {title}
  {\bibinfo {title} {Yield stress and elastic modulus of suspensions of
  noncolloidal particles in yield stress fluids},\ }\href@noop {} {\bibfield
  {journal} {\bibinfo  {journal} {Journal of Rheology}\ }\textbf {\bibinfo
  {volume} {52}},\ \bibinfo {pages} {287} (\bibinfo {year} {2008})}\BibitemShut
  {NoStop}%
\bibitem [{\citenamefont {Vasisht}\ \emph {et~al.}(2022)\citenamefont
  {Vasisht}, \citenamefont {Chaudhuri},\ and\ \citenamefont
  {Martens}}]{vasisht2022residual}%
  \BibitemOpen
  \bibfield  {author} {\bibinfo {author} {\bibfnamefont {V.~V.}\ \bibnamefont
  {Vasisht}}, \bibinfo {author} {\bibfnamefont {P.}~\bibnamefont {Chaudhuri}},\
  and\ \bibinfo {author} {\bibfnamefont {K.}~\bibnamefont {Martens}},\
  }\bibfield  {title} {\bibinfo {title} {Residual stress in athermal soft
  disordered solids: insights from microscopic and mesoscale models},\
  }\href@noop {} {\bibfield  {journal} {\bibinfo  {journal} {Soft Matter}\
  }\textbf {\bibinfo {volume} {18}},\ \bibinfo {pages} {6426} (\bibinfo {year}
  {2022})},\ \Eprint {https://arxiv.org/abs/2108.12782} {2108.12782}
  \BibitemShut {NoStop}%
\bibitem [{\citenamefont {Aime}\ \emph {et~al.}(2023)\citenamefont {Aime},
  \citenamefont {Truzzolillo}, \citenamefont {Pine}, \citenamefont {Ramos},\
  and\ \citenamefont {Cipelletti}}]{aimeUnifiedStateDiagram2023}%
  \BibitemOpen
  \bibfield  {author} {\bibinfo {author} {\bibfnamefont {S.}~\bibnamefont
  {Aime}}, \bibinfo {author} {\bibfnamefont {D.}~\bibnamefont {Truzzolillo}},
  \bibinfo {author} {\bibfnamefont {D.~J.}\ \bibnamefont {Pine}}, \bibinfo
  {author} {\bibfnamefont {L.}~\bibnamefont {Ramos}},\ and\ \bibinfo {author}
  {\bibfnamefont {L.}~\bibnamefont {Cipelletti}},\ }\bibfield  {title}
  {\bibinfo {title} {A unified state diagram for the yielding transition of
  soft colloids},\ }\href {https://doi.org/10.1038/s41567-023-02153-w}
  {\bibfield  {journal} {\bibinfo  {journal} {Nature Physics}\ }\textbf
  {\bibinfo {volume} {19}},\ \bibinfo {pages} {1673} (\bibinfo {year}
  {2023})}\BibitemShut {NoStop}%
\bibitem [{\citenamefont {Giesekus}(1982)}]{giesekus1982simple}%
  \BibitemOpen
  \bibfield  {author} {\bibinfo {author} {\bibfnamefont {H.}~\bibnamefont
  {Giesekus}},\ }\bibfield  {title} {\bibinfo {title} {A simple constitutive
  equation for polymer fluids based on the concept of deformation-dependent
  tensorial mobility},\ }\href@noop {} {\bibfield  {journal} {\bibinfo
  {journal} {Journal of Non-Newtonian Fluid Mechanics}\ }\textbf {\bibinfo
  {volume} {11}},\ \bibinfo {pages} {69} (\bibinfo {year} {1982})}\BibitemShut
  {NoStop}%
\bibitem [{\citenamefont {Ithaca}\ \emph {et~al.}(1986)\citenamefont {Ithaca},
  \citenamefont {Doi},\ and\ \citenamefont {Edwards}}]{ithaca1986theory}%
  \BibitemOpen
  \bibfield  {author} {\bibinfo {author} {\bibfnamefont {N.}~\bibnamefont
  {Ithaca}}, \bibinfo {author} {\bibfnamefont {M.}~\bibnamefont {Doi}},\ and\
  \bibinfo {author} {\bibfnamefont {S.}~\bibnamefont {Edwards}},\ }\href@noop
  {} {\bibinfo {title} {The theory of polymer dynamics}} (\bibinfo {year}
  {1986})\BibitemShut {NoStop}%
\bibitem [{\citenamefont {Cao}\ \emph {et~al.}(2023)\citenamefont {Cao},
  \citenamefont {Das}, \citenamefont {Windbacher}, \citenamefont {Ginot},
  \citenamefont {Kr{\"u}ger},\ and\ \citenamefont {Bechinger}}]{cao2023memory}%
  \BibitemOpen
  \bibfield  {author} {\bibinfo {author} {\bibfnamefont {X.}~\bibnamefont
  {Cao}}, \bibinfo {author} {\bibfnamefont {D.}~\bibnamefont {Das}}, \bibinfo
  {author} {\bibfnamefont {N.}~\bibnamefont {Windbacher}}, \bibinfo {author}
  {\bibfnamefont {F.}~\bibnamefont {Ginot}}, \bibinfo {author} {\bibfnamefont
  {M.}~\bibnamefont {Kr{\"u}ger}},\ and\ \bibinfo {author} {\bibfnamefont
  {C.}~\bibnamefont {Bechinger}},\ }\bibfield  {title} {\bibinfo {title}
  {Memory-induced magnus effect},\ }\href@noop {} {\bibfield  {journal}
  {\bibinfo  {journal} {Nature Physics}\ }\textbf {\bibinfo {volume} {19}},\
  \bibinfo {pages} {1904} (\bibinfo {year} {2023})}\BibitemShut {NoStop}%
\bibitem [{\citenamefont {Taylor}(1934)}]{taylor1934mechanism}%
  \BibitemOpen
  \bibfield  {author} {\bibinfo {author} {\bibfnamefont {G.~I.}\ \bibnamefont
  {Taylor}},\ }\bibfield  {title} {\bibinfo {title} {The mechanism of plastic
  deformation of crystals. part i.—theoretical},\ }\href@noop {} {\bibfield
  {journal} {\bibinfo  {journal} {Proceedings of the Royal Society of London.
  Series A, Containing Papers of a Mathematical and Physical Character}\
  }\textbf {\bibinfo {volume} {145}},\ \bibinfo {pages} {362} (\bibinfo {year}
  {1934})}\BibitemShut {NoStop}%
\bibitem [{\citenamefont {Swift}(1947)}]{swift1947elastic}%
  \BibitemOpen
  \bibfield  {author} {\bibinfo {author} {\bibfnamefont {H.}~\bibnamefont
  {Swift}},\ }\bibfield  {title} {\bibinfo {title} {Elastic deformation of
  piston rings},\ }\href@noop {} {\bibfield  {journal} {\bibinfo  {journal}
  {Engineering}\ }\textbf {\bibinfo {volume} {163}},\ \bibinfo {pages} {161}
  (\bibinfo {year} {1947})}\BibitemShut {NoStop}%
\bibitem [{\citenamefont {Molinari}\ \emph {et~al.}(1997)\citenamefont
  {Molinari}, \citenamefont {Ahzi},\ and\ \citenamefont
  {Kouddane}}]{molinari1997self}%
  \BibitemOpen
  \bibfield  {author} {\bibinfo {author} {\bibfnamefont {A.}~\bibnamefont
  {Molinari}}, \bibinfo {author} {\bibfnamefont {S.}~\bibnamefont {Ahzi}},\
  and\ \bibinfo {author} {\bibfnamefont {R.}~\bibnamefont {Kouddane}},\
  }\bibfield  {title} {\bibinfo {title} {On the self-consistent modeling of
  elastic-plastic behavior of polycrystals},\ }\href@noop {} {\bibfield
  {journal} {\bibinfo  {journal} {Mechanics of materials}\ }\textbf {\bibinfo
  {volume} {26}},\ \bibinfo {pages} {43} (\bibinfo {year} {1997})}\BibitemShut
  {NoStop}%
\bibitem [{\citenamefont {Sun}\ \emph {et~al.}(2016)\citenamefont {Sun},
  \citenamefont {Concustell}, \citenamefont {Carpenter}, \citenamefont {Qiao},
  \citenamefont {Rayment},\ and\ \citenamefont
  {Greer}}]{sunFlowinducedElasticAnisotropy2016}%
  \BibitemOpen
  \bibfield  {author} {\bibinfo {author} {\bibfnamefont {Y.~H.}\ \bibnamefont
  {Sun}}, \bibinfo {author} {\bibfnamefont {A.}~\bibnamefont {Concustell}},
  \bibinfo {author} {\bibfnamefont {M.~A.}\ \bibnamefont {Carpenter}}, \bibinfo
  {author} {\bibfnamefont {J.~C.}\ \bibnamefont {Qiao}}, \bibinfo {author}
  {\bibfnamefont {A.~W.}\ \bibnamefont {Rayment}},\ and\ \bibinfo {author}
  {\bibfnamefont {A.~L.}\ \bibnamefont {Greer}},\ }\bibfield  {title} {\bibinfo
  {title} {Flow-induced elastic anisotropy of metallic glasses},\ }\href
  {https://doi.org/10.1016/j.actamat.2016.04.022} {\bibfield  {journal}
  {\bibinfo  {journal} {Acta Materialia}\ }\textbf {\bibinfo {volume} {112}},\
  \bibinfo {pages} {132} (\bibinfo {year} {2016})}\BibitemShut {NoStop}%
\bibitem [{\citenamefont {Schott}\ \emph {et~al.}(2024)\citenamefont {Schott},
  \citenamefont {Dollet}, \citenamefont {Santucci}, \citenamefont
  {Schlep{\"u}tz}, \citenamefont {Claudet}, \citenamefont {Gst{\"o}hl},
  \citenamefont {Raufaste},\ and\ \citenamefont
  {Mokso}}]{schottMultiscaleStressDynamics2024}%
  \BibitemOpen
  \bibfield  {author} {\bibinfo {author} {\bibfnamefont {F.}~\bibnamefont
  {Schott}}, \bibinfo {author} {\bibfnamefont {B.}~\bibnamefont {Dollet}},
  \bibinfo {author} {\bibfnamefont {S.}~\bibnamefont {Santucci}}, \bibinfo
  {author} {\bibfnamefont {C.~M.}\ \bibnamefont {Schlep{\"u}tz}}, \bibinfo
  {author} {\bibfnamefont {C.}~\bibnamefont {Claudet}}, \bibinfo {author}
  {\bibfnamefont {S.}~\bibnamefont {Gst{\"o}hl}}, \bibinfo {author}
  {\bibfnamefont {C.}~\bibnamefont {Raufaste}},\ and\ \bibinfo {author}
  {\bibfnamefont {R.}~\bibnamefont {Mokso}},\ }\href
  {https://doi.org/10.48550/arXiv.2411.10338} {\bibinfo {title} {Multiscale
  stress dynamics in sheared liquid foams revealed by tomo-rheoscopy}}
  (\bibinfo {year} {2024}),\ \Eprint {https://arxiv.org/abs/2411.10338}
  {arXiv:2411.10338 [cond-mat]} \BibitemShut {NoStop}%
\bibitem [{\citenamefont {M{\'e}tivier}\ \emph {et~al.}(2012)\citenamefont
  {M{\'e}tivier}, \citenamefont {Rharbi}, \citenamefont {Magnin},\ and\
  \citenamefont {Bou~Abboud}}]{christel2012stick}%
  \BibitemOpen
  \bibfield  {author} {\bibinfo {author} {\bibfnamefont {C.}~\bibnamefont
  {M{\'e}tivier}}, \bibinfo {author} {\bibfnamefont {Y.}~\bibnamefont
  {Rharbi}}, \bibinfo {author} {\bibfnamefont {A.}~\bibnamefont {Magnin}},\
  and\ \bibinfo {author} {\bibfnamefont {A.}~\bibnamefont {Bou~Abboud}},\
  }\bibfield  {title} {\bibinfo {title} {Stick-slip control of the carbopol
  microgels on polymethyl methacrylate transparent smooth walls},\ }\href@noop
  {} {\bibfield  {journal} {\bibinfo  {journal} {Soft Matter}\ }\textbf
  {\bibinfo {volume} {8}},\ \bibinfo {pages} {7365} (\bibinfo {year}
  {2012})}\BibitemShut {NoStop}%
\bibitem [{\citenamefont
  {Risken}(1996)}]{riskenFokkerPlanckEquationMethods1996}%
  \BibitemOpen
  \bibfield  {author} {\bibinfo {author} {\bibfnamefont {H.}~\bibnamefont
  {Risken}},\ }\href@noop {} {\emph {\bibinfo {title} {The {{Fokker-Planck}}
  Equation: Methods of Solution and Applications}}},\ \bibinfo {edition}
  {second edition, 3rd printing}\ ed.,\ \bibinfo {series} {Springer Series in
  Synergetics}\ No.~\bibinfo {number} {18}\ (\bibinfo  {publisher} {Springer},\
  \bibinfo {address} {Berlin Heidelberg},\ \bibinfo {year} {1996})\BibitemShut
  {NoStop}%
\bibitem [{\citenamefont {Agoritsas}\ \emph {et~al.}(2015)\citenamefont
  {Agoritsas}, \citenamefont {Bertin}, \citenamefont {Martens},\ and\
  \citenamefont {Barrat}}]{agoritsasRelevanceDisorderAthermal2015}%
  \BibitemOpen
  \bibfield  {author} {\bibinfo {author} {\bibfnamefont {E.}~\bibnamefont
  {Agoritsas}}, \bibinfo {author} {\bibfnamefont {E.}~\bibnamefont {Bertin}},
  \bibinfo {author} {\bibfnamefont {K.}~\bibnamefont {Martens}},\ and\ \bibinfo
  {author} {\bibfnamefont {J.-L.}\ \bibnamefont {Barrat}},\ }\bibfield  {title}
  {\bibinfo {title} {On the relevance of disorder in athermal amorphous
  materials under shear},\ }\href {https://doi.org/10.1140/epje/i2015-15071-x}
  {\bibfield  {journal} {\bibinfo  {journal} {The European Physical Journal E}\
  }\textbf {\bibinfo {volume} {38}},\ \bibinfo {pages} {71} (\bibinfo {year}
  {2015})},\ \Eprint {https://arxiv.org/abs/1501.04515} {1501.04515}
  \BibitemShut {NoStop}%
\bibitem [{\citenamefont {Olivier}\ and\ \citenamefont
  {Renardy}(2013)}]{olivierGeneralizationHebraudLequeux2013}%
  \BibitemOpen
  \bibfield  {author} {\bibinfo {author} {\bibfnamefont {J.}~\bibnamefont
  {Olivier}}\ and\ \bibinfo {author} {\bibfnamefont {M.}~\bibnamefont
  {Renardy}},\ }\bibfield  {title} {\bibinfo {title} {On the {{Generalization}}
  of the {{H{\'e}braud}}--{{Lequeux Model}} to {{Multidimensional Flows}}},\
  }\href {https://doi.org/10.1007/s00205-012-0603-7} {\bibfield  {journal}
  {\bibinfo  {journal} {Archive for Rational Mechanics and Analysis}\ }\textbf
  {\bibinfo {volume} {208}},\ \bibinfo {pages} {569} (\bibinfo {year}
  {2013})}\BibitemShut {NoStop}%
\end{thebibliography}

%

\end{document}


\title{Supplementary information for Disorder-induced stress-flow misalignment in soft glassy materials revealed using multi-directional shear}

\author{Frédéric Blanc}
\affiliation{Institut de Physique de Nice, UMR7010, CNRS, Université Côte d'Azur, 17 rue Julien Lauprêtre, 06200 Nice, France}
\author{Guillaume Ovarlez}
\affiliation{Univ. Bordeaux, CNRS, Syensqo, LOF, UMR 5258, F-33600 Pessac, France}
\author{Adam Trigui}
\author{Kirsten Martens}
\author{Romain Mari}
\affiliation{Univ. Grenoble-Alpes, CNRS, LIPhy, 38000 Grenoble, France}


\maketitle


\section*{Stress-flow misalignement in the kinematic hardening model}

\subsection*{Kinematic hardening model}

We consider a YSF under deformation. 
If a spatial point of position $\bm{x}$ at time $t$ was at position $\bm{X}$ before the deformation, we can define deformation gradient $\bm{F} = \nabla_{\bm{X}} \bm{x}(t)$.
The elasto-viscoplastic model of Dimitriou \& McKinley~\cite{dimitriou2013describing,dimitriou2019canonical} is based on the assumption of multiplicative decomposition of the deformation gradient in elastic and plastic parts~\cite{gurtinMechanicsThermodynamicsContinua},
\begin{equation}
    \bm{F} = \bm{F}^\mathrm{e}\bm{F}^\mathrm{p}\ .
\end{equation}

The elastic-plastic decomposition of the deformation gradient allows to consider a structural space which is the image of the undeformed material by application of the plastic deformation $\bm{F}^\mathrm{p}$ only.

We can also define a velocity gradient $\bm{L} = \nabla_{\bm{x}} \dot{\bm{x}}$, which satisfies an additive decomposition in elastic and plastic parts
\begin{equation}
    \bm{L} = \bm{L}^\mathrm{e} + \bm{F}^\mathrm{e}\bm{L}^\mathrm{p}{\bm{F}^\mathrm{e}}^{-1}\, .
\end{equation}
The model assumes that the plastic velocity gradient is irrotational, $\bm{L}^\mathrm{p} = {\bm{L}^\mathrm{p}}^T$ and volume-preserving, $\text{tr}(\bm{L}^\mathrm{p})=0$. 

The Cauchy stress of the material $\bm{\Sigma}$ is most easily expressed using the second Piola stress tensor (which is the image of the Cauchy stress through application of ${\bm{F}^\mathrm{e}}^{-1}$, that is, in the structural space), $\bm{\Sigma}^\mathrm{e} = J {\bm{F}^\mathrm{e}}^{-1} \bm{\Sigma}{\bm{F}^\mathrm{e}}^{-T}$, with $J = \det{\bm{F}^\mathrm{e}}$.
The stress is of purely elastic nature
\begin{equation}
    \bm{\Sigma}^\mathrm{e} = 2 G \bm{D}^\mathrm{e} + \Lambda (\text{tr} \bm{D}^\mathrm{e})\bm{1}\, ,
\end{equation}
with $\bm{D}^\mathrm{e}$ the Green-St. Venant strain tensor,
\begin{equation}
    \bm{D}^\mathrm{e} = \frac{1}{2}\left({\bm{F}^\mathrm{e}}^{T}\bm{F}^\mathrm{e} - \bm{1}\right)\, .
\end{equation}

The rest of the modeling amounts to prescribe a flow rule (a stress-strain rate relation) to close the model, that is, to be able to predict $\bm{L}^\mathrm{p}$ given the current stress state, or vice-versa.

The kinematic hardening is implemented in this structural space, and is assumed to be controlled there by a strain-like symmetric tensor field $\bm{A}$ with $\det \bm{A} = 1$~\cite{anandThermomechanicallyCoupledTheory2009}.
This structural tensor sets the so-called back stress tensor $\bm{M}_\mathrm{back} = C\log{\bm{A}}$, with $C$ a parameter called the back stress modulus. Note that $\bm{M}_\mathrm{back}$ is deviatoric.
The tensor $\bm{A}$ is given the dynamics~\cite{amesThermomechanicallyCoupledTheory2009}
\begin{equation}
    \dot{\bm{A}} = \bm{L}^\mathrm{p}\bm{A}+\bm{A}\bm{L}^\mathrm{p} - \sqrt{2} q  \bm{A}\log\bm{A} \lvert \bm{L}^\mathrm{p} \rvert\, ,
\end{equation}
with $q$ a inverse-strain-like parameter which together with the modulus $C$ controls the yield stress, as briefly discussed below.

In turn, the back stress controls the plasticity in the system. 
For this we introduce the Mandel stress, a symmetric tensor which is the conjugate of the plastic velocity gradient $\bm{L}^\mathrm{p}$ in the expression of plastic dissipation, as $\bm{M} = J {\bm{F}^\mathrm{e}}^T \bm{\Sigma} {\bm{F}^\mathrm{e}}^{-T}$, and its deviatoric part $\bm{M}_0 = \bm{M} - \frac{\text{tr}\bm{M}}{3}\bm{1}$.
The plastic flow is co-directional with the effective stress $\bm{M}_\mathrm{eff} = \bm{M}_0 - \bm{M}_\mathrm{back}$~\cite{anandThermomechanicallyCoupledTheory2009}, with a flow rule~\cite{dimitriou2019canonical}
\begin{equation}
    \bm{L}^\mathrm{p} = \frac{1}{\sqrt{2}}\left(\frac{\lvert \bm{M}_\mathrm{eff}\rvert}{\sqrt{2}k}\right)^{1/m}\frac{\bm{M}_\mathrm{eff}}{\lvert \bm{M}_\mathrm{eff}\rvert}\, .
\end{equation}
The specifics of the prefactors and exponents are here to ensure a Herschel-Bulkley flow curve with flow index $m$, consistency parameter $k$ and yield stress $\sim C/q$~\cite{dimitriou2019canonical}.

This flow rule closes the model. For a given prescribed velocity gradient $\bm{L}$ and a given stress $\bm{\Sigma}$, one can compute the plastic strain rate tensor $\bm{L}^\mathrm{p}$ and integrate the evolution for $\bm{A}$. 
The initialization of the model is $\bm{\Sigma}= \bm{A} = \bm{F}^\mathrm{e} = \bm{F}^\mathrm{p} = \bm{1}$.

\subsection*{Model predictions}

We here present the predictions of the kinematic hardening model for parameters that correspond to the steady-state rheology measured in the experiment, namely $m= 0.34$, $C = \SI{1206}{\pascal}$, $q = 10$, $k = \SI{57}{\pascal}\text{s}^m$, and elastic moduli $G=\SI{626}{\pascal}$ and $\Lambda = \infty$ (incompressible material).
We follow the experimental protocol (see main text) for the kinematic hardening model.
After applying a simple shear $\bm{L}=\dot\gamma \bm{e}_1\bm{e}_2$ until steady state is reached, we perform a shear cessation and let the stress tensor relax, before performing a quasistatic unloading to remove any residual stress.
By this point, the stress $\bm{\Sigma}$ vanishes, 
but the back stress $\bm{M}_\mathrm{back}$ is finite, as the pre-shear induced kinematic hardening.

From this rest state, we rotate $\bm{e}_1$ and $\bm{e}_3$ by an angle $\theta$ around the gradient direction $\bm{e}_2$, and apply a simple shear in the new direction, $\bm{L}=\dot\gamma \bm{e}_1\bm{e}_2$.
The resulting stresses $\Sigma^\theta_\parallel$ and $\Sigma^\theta_\perp$ are shown in Fig.~\ref{fig:Dimitriou}a and Fig.~\ref{fig:Dimitriou}b, respectively, with the same polar representation as for the experiments in the main text.

And indeed, the kinematic hardening model is able to recover the existence of a finite orthogonal stress $\Sigma^\theta_\perp$, with the right sign.
It however underestimates its value, predicting $|\Sigma^\theta_\perp|/\Sigma^\mathrm{ss}_\parallel \lesssim 0.03$.
It also underestimates the relaxation strain, as the orthogonal stress relaxes on a strain of order 1, just like the parallel stress, when in experiments we observe that the relaxation of $\Sigma^\theta_\perp$ occurs on strains of order 10, much large than that of $\Sigma^\theta_\parallel$.

Unsurprisingly, the model also shows kinematic hardening, as seen from the value at which the parallel stress reaches \SI{90}{\percent} of its steady-state value [black curve in Fig.~\ref{fig:Dimitriou}(a)]. 
While we observe some anisotropy, it is much weaker than in the experiments shown in Fig.~3(a) of the main text. 

In the model, the orthogonal stress stems from the fact that just after the shear rotation, the plastic part of the velocity gradient $\bm{F}^\mathrm{e}\bm{L}^\mathrm{p}{\bm{F}^\mathrm{e}}^{-1}$, which directly inherit its principal directions from the one of the pre-shear via the structural tensor $\bm{A}$, is not co-directional with the total velocity gradient $\bm{L}$.
This forces the elastic velocity gradient to compensate for this misalignment, which builds up an elastic deformation gradient with a finite $\bm{e}_3\bm{e}_2$ component, and therefore an orthogonal stress.
Therefore, in this model kinematic hardening and orthogonal stresses are two sides of the same coin, both stemming from the back stress.


\begin{figure}
    \centering
    \includegraphics[width=0.6\linewidth]{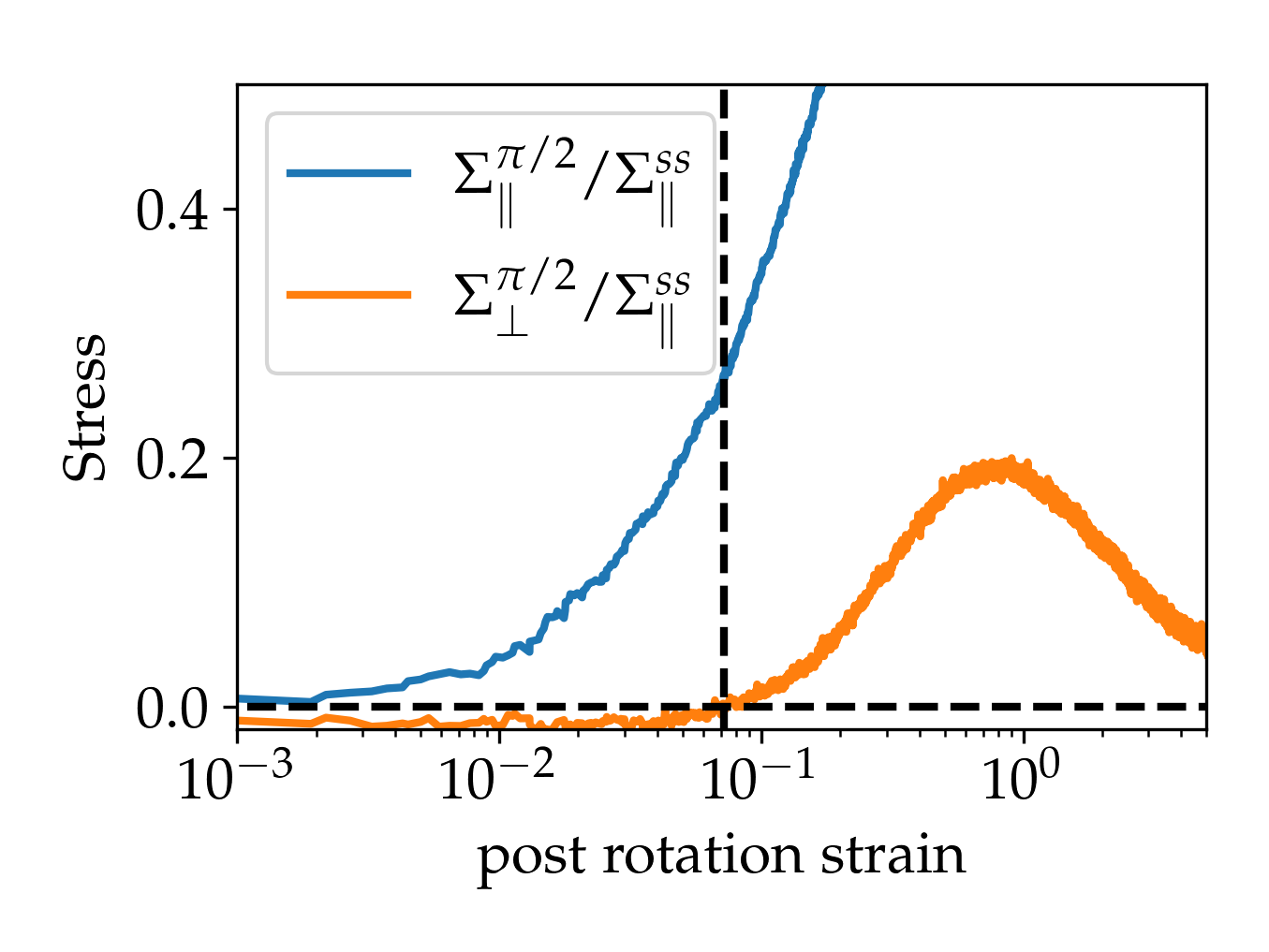}
\includegraphics[width=0.6\linewidth]{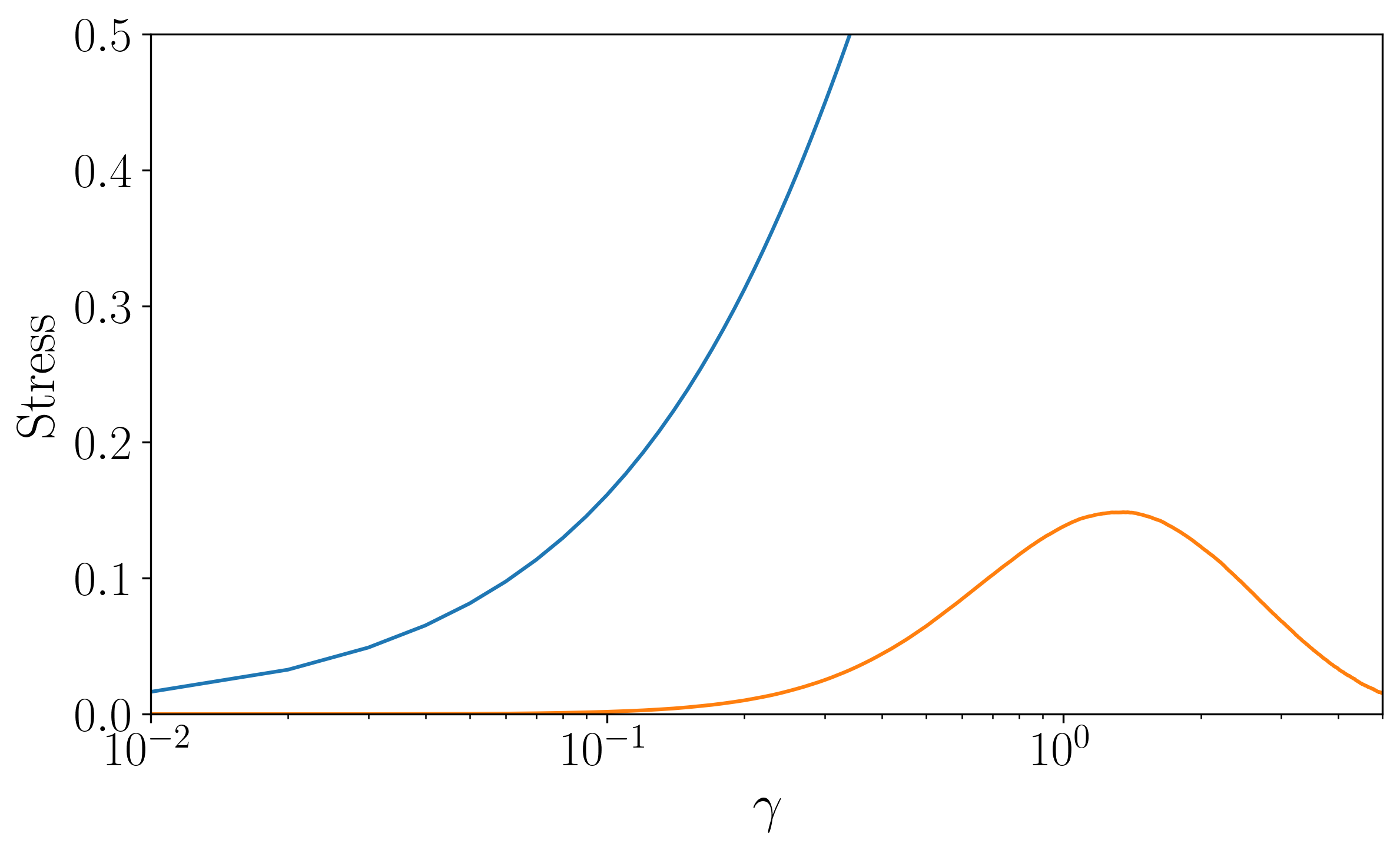}
    \caption{(a) Two-dimensional response of a Carbopol gel to a change of the shear direction in the 2D rheometer of Fig.~1 in the manuscript; flow is rotated by an angle $\theta=\pi/2$ with respect to the preshear direction. Dimensionless shear stress $\Sigma^{\theta}_\parallel/\Sigma^\mathrm{ss}_\parallel$ aligned with the shear direction (blue line) and dimensionless shear stress $\Sigma^{\theta}_\perp/\Sigma^\mathrm{ss}_\parallel$ orthogonal to the shear direction (orange line). The vertical dashed line show the strain where a first detectable change in the value of $\Sigma^{\theta}_\perp$ is observed. (b) Same plot for the model.}
    \label{fig:onsetofplasticityF32}
\end{figure}

\begin{figure}
    \centering
    \includegraphics[width=\linewidth]{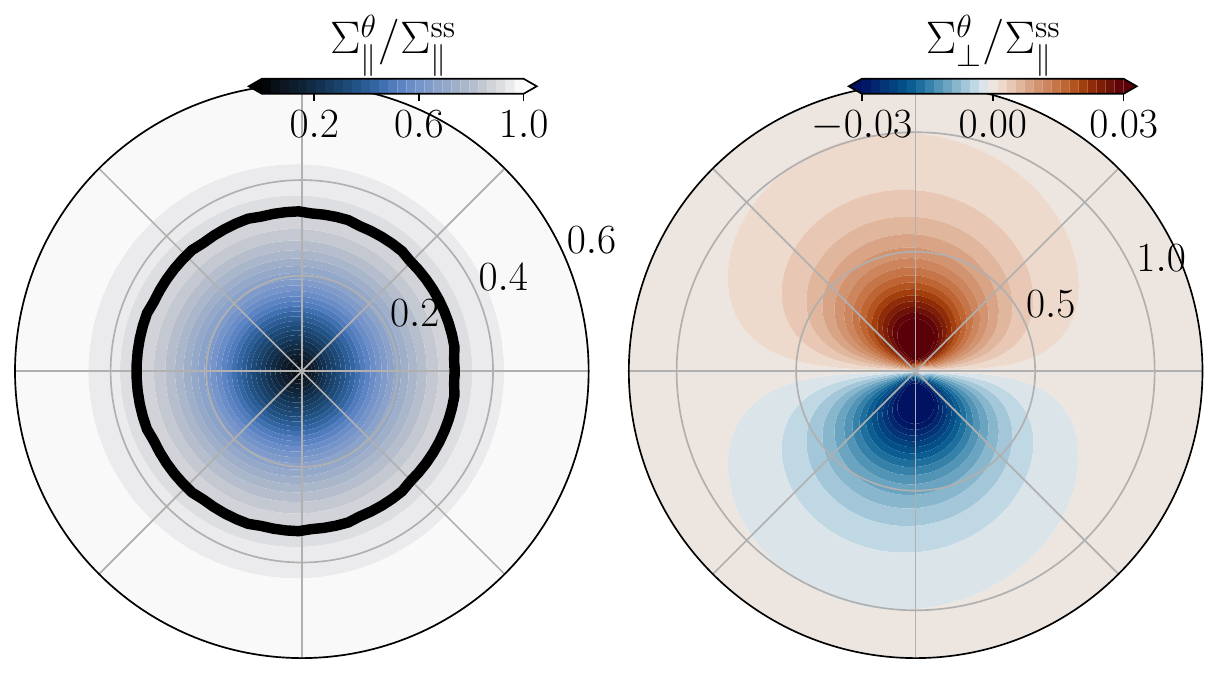}
    \caption{\footnotesize Predictions of the kinematic hardening model by Dimitriou \& McKinley~\cite{dimitriou2013describing,dimitriou2019canonical}. (a) Colormap of the dimensionless shear stress $\Sigma^{\theta}_\parallel/\Sigma^\mathrm{ss}_\parallel$ aligned with the shear direction. The radial coordinate is the strain applied after shear rotation and the azimuthal coordinate is the shear rotation angle $\theta$. The black contour marks the level $\Sigma^{\theta}_\parallel = 0.9\Sigma^\mathrm{ss}_\parallel$.(b) Dimensionless shear stress $\Sigma^{\theta}_\perp/\Sigma^\mathrm{ss}_\parallel$ orthogonal to the shear direction, in the same representation as panel (a). }
    \label{fig:Dimitriou}
\end{figure}



%